\documentclass[smallcondensed,onecolumn,a4]{svjour3}

\usepackage{graphicx}
\usepackage{footnote}
\usepackage{pdflscape}
\usepackage{natbib}
\usepackage{authblk}
\usepackage{tabu}
\usepackage{arydshln}
\usepackage{subscript}
\usepackage{url}
\usepackage{supertabular}
\usepackage{longtable}
\usepackage{multicol}
\usepackage{multirow}
\usepackage{amsmath}
\usepackage{breqn}
\usepackage{makecell}
\usepackage{diagbox}
\usepackage{booktabs} 
\usepackage{bbold}

\usepackage{calrsfs}
\usepackage{color}
\usepackage{bm}
\usepackage{changes}
\usepackage[T1]{fontenc}
\usepackage{ae,aecompl}

\smartqed  
\newcommand{\obf}[1][]{}

\newcommand{\finalstate}[1][]{%
  \renewcommand{\added}[2][]{{##2}}%
  \renewcommand{\replaced}[3][]{{##2}}%
  \renewcommand{\deleted}[2][]{}}

\definechangesauthor[name={Pierre}]{P}
\definechangesauthor[name={Agnes}, color=teal]{A}
\definechangesauthor[name={Ashok}, color=magenta]{V}
\begin{document}
\finalstate

\title{Gaia-DR2 asteroid observations and INPOP planetary ephemerides}

\author{
P. Deram$^{1}$
\and A. Fienga$^{1,2}$
\and A. K. Verma$^{3}$
\and M. Gastineau$^{2}$
\and J. Laskar$^{2}$
 }
 
\titlerunning{Gaia-DR2 asteroid observations and INPOP}
\authorrunning{Deram et al.}

\institute{
G\'eoAzur, CNRS, Observatoire de la C\^ote d'Azur, Universit\'e C\^ote d'Azur, 250 Av. A. Einstein, Valbonne, 06560, France
\and IMCCE, Observatoire de Paris, PSL University, CNRS, Sorbonne Universit\'e, 77 avenue Denfert-Rochereau, Paris, 75014, France
\and  University \added[id=V]{of} California, Los Angeles, CA, USA
}

%
%
\date{Received: date / Accepted: date}

\maketitle

\begin{abstract}
   
 We {\obf{used}} the \deleted[id=A]{\replaced[id=P]{newly}{new}  released} INPOP19a planetary ephemerides to perform the orbital adjustment of 14099 asteroids based on \deleted[id=P]{the} Gaia-DR2 observations, and compare \added[id=A]{for 23 of them} \replaced[id=P]{the resulting orbits}{them}  to radar data. As Gaia-DR2 has been processed using the planetary \replaced[id=V]{ephemeris}{ephemerides} INPOP10e, \replaced[id=V]{the primary goal}{one of the goal} of this paper is to confirm the portability of the data when using an updated version of the \added[id=A]{solar} system model. \replaced[id=P]{In particular} {Especially}, we point \added[id=P]{out} the fact that the Gaia \replaced[id=A]{satellite}{spacecraft} position\added[id=A]{s} - provided with respect to the INPOP10e \added[id=A]{solar} system barycenter - must be corrected when using another \replaced[id=P]{planetary ephemeris}{ephemerides} . We also present a convenient least square formalism that only handles small matrice\added[id=P]{s} and allows the adjustment of global parameters\added[id=A]{, such as masses}. In order to \replaced[id=A]{check}{verify} the \replaced[id=A]{consistency}{portability} of the Gaia observations \added[id=A]{with other types of observations}, we perform \added[id=A]{an} orbital adjustment \replaced[id=A]{in combining}{with combined} Gaia \replaced[id=P]{and radar {\obf{range}} observations for 23 objects, \replaced[id=A]{together} {completed} with a careful post-fit analysis including an estimation of the Gaia systematic errors.}{,radar and ground-based observations for 23 objects using two different methods: a systematic exploration of the weighting scheme coupled with a residual post-fit analysis, and a Least Square Variance Component Estimation algorithm (LSVCE) } \added[id=A]{Finally, we show that to ensure the combined use of Gaia angular DR2 observations and radar ranging, a more developed than firstly proposed dynamical modeling is required together with the addition of the systematic Gaia bias in the fit procedure.These results give promising directions for the next Gaia delivery, Gaia-DR3.} \\
 \end{abstract}
 
 \section{\replaced[id=P]{Asteroid observations in Gaia-DR2}{Asteroids Gaia-DR2} }

In 2013, the \deleted[id=P]{astrometric} G\replaced[id=P]{aia}{AIA} \replaced[id=A]{astrometric satellite}{spacecraft} of the European Space Agency (ESA)  was launched  at the \deleted[id=A]{second} Lagrange point L2 \replaced[id=A]{for}{with the main mission of} surveying \deleted[id=V]{with an unprecedented \added[id=P]{astrometric} accuracy (down to 24 microarcseconds)} the stellar population of the Milky way \citep{2016A&A...595A...1G} \added[id=V]{with an unparalleled precision of astrometric accuracy (down to 24 \textmu arcseconds) }. Parallel to this main mission \replaced[id=A]{,}{-} representing a billion of observations \replaced[id=A]{,}{-}  the \replaced[id=A]{Gaia capability}{characteristic of Gaia} \replaced[id=A]{of observing}{, especially for} small \replaced[id=V]{extended}{extented} target\added[id=A]{s} was also employed to provide a large survey of detected Solar System objects. 

In 2018, as part of the \replaced[id=A]{Gaia Data Release 2 (Gaia-DR2)}{Gaia-DR2}, \replaced[id=A]{were}{the mission} released \deleted[id=P]{for the first time} the positions \deleted[id=P]{and velocities} of about 14 099 known \replaced[id=A]{asteroids}{Solar System objects} based on 1 977 702 observations acquired during \added[id=A]{the first} 22 months of \added[id=A]{the} mission between August 5,2014 and May 23, 2016 \citep{refId0}. \replaced[id=P]{Gaia-DR2}{it}  includes mainly asteroids from the \replaced[id=A]{Main Belt}{main belt} and  Near-Earth\added[id=V]{,} and Kuiper belt  objects \deleted[id=P]{(Trojans  and 2 TNOs)}. The \added[id=V]{archived} positions \replaced[id=A]{are given}{were represented} in the G\replaced[id=P]{aia}{AIA} specific coordinates AL and AC (respectively along and across the Gaia scan direction\added[id=A]{s})\added[id=V]{,} with an optimal range of brightness G \replaced[id=A]{between 12 and 17}{$=$12-17} where the accuracy in the AL direction reaches \added[id=A]{the} milliarcsecond \added[id=A]{level (mas)}. As the error\added[id=A]{s} \replaced[id=V]{in the  AC direction is}{on AC remain\deleted[id=P]{s}} considerably larger, the information provided by G\replaced[id=P]{aia}{AIA} is essentially 1D. However, due to the large variety of orientations and scan directions covered by Gaia over time (a minimal number of 12 transits per object\deleted[id=V]{s} \replaced[id=V]{were}{was} requested for DR2), \replaced[id=P]{the AL direction alone provide\added[id=A]{s} strong orbital constraints}{it is sufficient to constraint the orbit}\added[id=V]{ \citep{2018A&Atest}}. 

\cite{2018A&Atest} presented the main step of the \added[id=A]{solar system objects} data processing and performed a first orbital \replaced[id=A]{fit}{fitting procedure} based \deleted[id=V]{only} on Gaia observations \added[id=V]{only}. \replaced[id=V]{They included 16 massive perturbers in their dynamical model and used JPL’s DE431 planetary ephemeris \citep{2014IPNPR.196C...1F} to access the positions of the planets }{This adjustment was obtained using a dynamical model \replaced[id=A]{including}{with} 16 perturbing massive bodies, and \added[id=A]{the} planet orbits \replaced[id=A]{were taken from the JPL DE431 planetary ephemeris \citep{2014IPNPR.196C...1F}}{provided by DE431 ephemeris \citep{2014IPNPR.196C...1F}}}. In the AL direction,  \replaced[id=V]{they reported}{\replaced[id=P]{the orbital fit \deleted[id=P]{s} give\added[id=A]{s}}{it leads to}} a mean residual of 0.05 mas with a standard deviation of 2.14 mas \added[id=V]{which is} in accordance with the announced accuracy. For object\added[id=P]{s} brighter than G=13, the same order of \replaced[id=A]{accuracy}{precision} \replaced[id=V]{was also reported}{is also reached} in the AC direction as a full 2D window \replaced[id=V]{was}{is} transmitted (with a larger pixel size in AC). However, as expected, for objects with G>13, the error in AC r\replaced[id=V]{remained}{remains} considerably larger (of \replaced[id=A]{about}{the order of} 600 mas) as all pixel\added[id=A]{s} \replaced[id=V]{were}{are} binned to a single window. 

\replaced{These}{Those} promising results \replaced[id=V]{were}{,} based \replaced[id=A]{only on}{on only} 22 month\added[id=P]{s} of observations\replaced[id=V]{ and pose}{, open} great opportunities \deleted[id=A]{for study}\added[id=V]{for future research}\deleted[id=V]{,} as a \deleted[id=P]{future} release of nearly 100 000 objects is expected \deleted[id=P]{in the future} \added[id=A]{ with the Gaia-DR3}, \replaced[id=A]{and a time coverage of 5 years}{covering \deleted[id=P]{a} \replaced[id=A]{an almost 5-years}{longer} time-span}. 
However, as the Gaia\added[id=A]{-DR2} observations were processed using the planetary ephemerides INPOP10e \citep{INPOP10e,2013arXiv1301.1510F} \added[id=A]{and as planetary ephemerides have {\obf{evolved}} since 2013} , it is worth \replaced[id=P]{checking}{to verify} the {\obf{compatibility}} of \replaced[id=V]{these observations}{such data} when \replaced[id=A]{using}{working with} an updated version of Solar system ephemeris, \replaced[id=A]{for investigating mainly}{in order to study} the potential impact of \deleted[id=P]{the} model \replaced[id=P]{dependencies}{dependence} and the  portability of the released observations. 

This \added[id=A]{planetary ephemeris} model dependenc\replaced[id=A]{y}{e} operates at different steps and levels during the processing of the Gaia observations \citep{2016NSTIM.104.....F}.
A prior knowledge of the position and velocity vectors of the \replaced[id=A]{satellite}{spacecraft} with respect to the barycenter of the \replaced[id=P]{S}{s}olar system \added[id=A]{(SSB)} is for example necessary in order to \replaced[id=A]{apply the aberration corrections to}{correct} the source apparent positions \deleted[id=A]{for the aberration}\replaced[id=V]{ and }{. The distance to the \replaced[id=A]{SSB}{\replaced[id=P]{S}{s}olar system barycenter} is also needed} to determine the stellar parallaxes with the proper scale. 
\replaced[id=V]{It is also required to process the astrometric data}{Finally, it is also required that the astrometric data processing was carried out} in\added[id=A]{to} \deleted[id=V]{the} BCRS (Barycentric celestial reference System) with \replaced[id=A]{its}{the} origin at the \replaced[id=A]{SSB}{barycenter of the \replaced[id=P]{S}{s}olar system}, and \added[id=A]{its} reference direction\added[id=A]{s} \replaced[id=A]{tied to}{provided by} the International Celestial Reference System (ICRS). 

The goal of this paper is first to \replaced[id=A]{check}{verify} the compatibility of the Gaia observations processed with INPOP10e, when we  \replaced[id=A]{adjust}{try an adjustment} with the latest version of INPOP. Second\added[id=V]{ly}, \added[id=P]{we seek} to validate our adjustment method in anticipation of the huge amount of observations that will be available with \deleted[id=P]{the} DR3 \added[id=A]{by confronting estimated orbits to very accurate radar observations.}. \deleted[id=P]{Finally, we aim to improve the accuracy of the INPOP planetary ephemerides, especially concerning Mars orbit which is the most sensitive to the SSOs (Solar System Objects) dynamic and mass.} 

In Sect. \ref{ephemeride}, we provide an overview of the differences between INPOP10e and INPOP19a ephemerides \added[id=P]{\citep{2019NSTIM.109.....V,2019MNRAS.tmp.3035F}}. The details of the inversion method used in this study are discussed in Sect. \ref{adjustment}. \added[id=P]{The result\added[id=V]{s} of the joint analysis that includes \added[id=A]{Gaia} \replaced[id=A]{satellite}{spacecraft} and ground-based radar data are provided in Sect. \ref{postfit}\replaced[id=A]{. In Sect. \ref{solution}, w}{ and \ref{solution} where w}e analyse the impact of the weighting schema and perturber\deleted[id=P]{s} list on the obtained residuals.}

\section{Planetary ephemerides}

\label{ephemeride}

\replaced[id=V]{The INPOP planetary ephemeris (Intégrateur Numérique Planétaire de l'Observatoire de Paris) was built in 2006  \citep {fienga:2008aa} and has become a reference for science in solar system dynamics and fundamental physics.}{INPOP (Int\'egrateur Num\'erique Plan\'etaire de l'Observatoire de Paris) is a planetary ephe\-meris created in 200\replaced[id=A]{3}{6} \citep {fienga:2008aa} \replaced[id=P]{that}{which}  has become a reference for scientific research in dynamics of the \replaced[id=P]{S}{s}olar system objects and in fundamental physics}.INPOP numerically integrates \replaced[id=A]{the orbits of}{the solar system bodies that include planets, the Moon and asteroids} \added[id=P]{the Sun, planets, the Moon\added[id=V]{,} and selected asteroids}. The position\added[id=A]{s} and velocit\added[id=A]{ies} of these bodies are \replaced[id=V]{estimated}{adjusted} using more than 150,000 observations such as lunar laser ranging, spacecraft \replaced[id=A]{tracking}{range}, and ground-based observations. \replaced[id=A]{The INPOP ephemerides}{The positions and velocities of the objects} are regularly updated following the constant improvement of the Solar system model and the addition of new observations\replaced[id=A]{. They are}{, and} distributed at \url{http://www.imcce.fr/inpop}. \citep{fienga:2008aa,fienga:2011cm,fienga:2015cm,fienga:2019inpop}. \added[id=P]{We present} two ephemerides \deleted[id=P]{were presented} in the following subsection\added[id=P]{s}: INPOP10e, used in the Gaia data processing, and INPOP19a, the latest version used in this paper for the \added[id=A]{asteroid} orbital adjustment.

\begin{table}[]
    \caption{Differences between INPOP10e and INPOP19a planetary ephemerides. \replaced[id=V]{The first part of the table provides a breakdown of the differences based on dynamic modeling and adjustment. The second part of the table summarizes additional data added since INPOP10e that was used to construct INPOP19a.}{In a first part, two categories are considered: the differences in the dynamical modeling and in the adjustement \deleted[id=P]{process}. \replaced[id=P]{The second part of table provides}{In a second part, are given}  the datasets used for the construction of INPOP19a and added since the publication of INPOP10e. }{\obf{ More details can be found in Sect \ref{inpop10e} and \ref{inpop19a}}}}
    \centering
    \begin{tabular}{l|c c}
        \hline
    & INPOP10e & INPOP19a \\
    & \cite{INPOP10e, 2013arXiv1301.1510F} & \cite{2019MNRAS.tmp.3035F,fienga:2019inpop}\\
    & & \cite{DiRuscio2020} \\
    \hline
    {\bf{Dynamical modeling}} & & \\
         Main belt asteroid perturbations &  343 $+$ 1 ring & 343 \\
         TNO perturbations & none & 3 rings \\ 
         Moon libration and core rotation & approximated for fluid core,  & full for fluid core,\\
         & no inner core & no inner core \\
    {\bf{Fit and estimated parameters}} & & \\
        total number of estimated parameters & 210 & 402  \\
         asteroid masses & 152 & 343 \\
         ring mass & main belt & TNO \\
     Dataset Time interval & 1913:2010 & 1924:2017 \\
     \hline
     \hline
     {\bf{INPOP19a Additional datasets}} & Planets & Time coverage \\
     \hline
        Messenger & Mercury & 2011:2014 \\
        Cassini & Saturn & 2004:2017 \\
        Juno & Jupiter & 2016:2019 \\
        MEX & Mars & 2014:2017 \\
        MRO & Mars & 2010:2014 \\
        \hline
    \end{tabular}
    \label{tab:10ev19a}
\end{table}

\subsection{INPOP10e} 
\label{inpop10e}
A first version of INPOP, INPOP06, was published in 2008 \citep{2008jsrs.meet...69F} with a  dynami\added[id=A]{cal} model and a fit procedure very close to the  JPL reference ephemerides of the time. \replaced[id=V]{The addition of new observational data and the evolution of the dynamical models and adjustment techniques led to the successive release of the INPOP series, followed by INPOP08 \citep{2009A&A...507.1675F}, INPOP10a \citep{2011CeMDA.111..363F}, and INPOP10e \citep{2013arXiv1301.1510F}, developed for the Gaia mission and released in 2013.}{The addition of new observational data, the development of the Moon and planetary model\added[id=P]{s} and new adjustment methods \added[id=P]{led}\deleted[id=P]{leads}  to the construction of successively INPOP08 \citep{2009A&A...507.1675F}, INPOP10a \citep{2011CeMDA.111..363F} and finally INPOP10e \citep{2013arXiv1301.1510F}, developed for the Gaia mission\deleted[id=P]{,} and released in 2013.} 

\added[id=A]{Besides providing the user} \deleted[id=P]{INPOP provides the user} \added[id=A]{with} positions and velocities of the planets \added[id=A]{and} the Moon, \added[id=A]{INPOP provides also} the rotation angles of the Earth and the Moon as well as TT-TDB in accordance with the recommendations of the \replaced[id=V]{I}{i}nternational Astronomical Union (IAU) in terms of time scale, metric (relativistic equation of motion)\added[id=V]{,} and \replaced[id=V]{S}{s}un gravitational mass \deleted[id=A]{\added[id=P]{and J2} adjustment} (\deleted[id=P]{adjustment of J2 oblateness} with a fixed \replaced[id=A]{astronomical unit (au)}{\added[id=P]{au}\deleted[id=P]{AU}} ). \deleted[id=P]{In INPOP10e, the Moon libration parameters, Moon and Earth potential coefficients and Earth-Moon barycenter mass were fitted with respect to the LLR observations.} \added[id=A]{ In INPOP10e,} 152 asteroid masses were also estimated using a sophisticated procedure based on a bounded least square associated with prior sigma and solar plasma correction\added[id=A]{s} \citep{2013A&A...550A.124V}. 

In accordance with the \added[id=P]{Gaia} mission \added[id=A]{and IAU} requirement\added[id=A]{s}, the link with the ICRF was maintained by the use of VLBI differential observations of \added[id=A]{planetary} spacecraft\added[id=A]{s} relative to \added[id=V]{the} ICRF source\added[id=A]{s}. The combination of such VLBI observations with spacecraft navigation provides the planet positions relative\deleted[id=P]{ly} to \added[id=V]{the} ICRF source\added[id=A]{s} at the \added[id=V]{level of} accuracy of the VLBI space mission localization (typically milliarcsecond). With the INPOP10e version, such \added[id=A]{a} tie was sustained by VLBI tracking data of inner planets and outer planets - especially \deleted[id=V]{the} Mars, VEX\added[id=V]{,} and Cassini observations - with an accuracy of about 1 mas \deleted[id=V]{over the past 10 years until 2013}\added[id=P]{for the period 2003-2013}.

In term\added[id=P]{s} of observations, the INPOP10e was the first INPOP version to  benefit from flyby observations in addition to the classical ground-based optical ones. For  the  inner  planets,  MEX  and  VEX \replaced[id=V]{radiometric}{radar tracking} and VLBI \replaced[id=V]{observations were crucial for}{have an important impact for the quality of the} Mars and Venus orbits. Uranus  and  Neptune \added[id=V]{orbits were}  also  benefited  from  the  use  of  positions  deduced from the Voyager 2 (two flyby\added[id=A]{s}). For \added[id=P]{M}\deleted[id=P]{m}ercury, three positions were \added[id=A]{deduced from} \deleted[id=P]{given by} the Messenger flybys while for Jupiter, five Juno flybys were used \deleted[id=A]{(Cassini)}. 

Typically the maximum difference between \replaced[id=V]{INPOP10e}{this version of INPOP} and \deleted[id=P]{and the JPL DE423} \added[id=P]{the \added[id=A]{contemporary} JPL\added[id=V]{'s} \added[id=A]{ephemeris} \deleted[id=P]{available at the same time} - \added[id=A]{DE420}\deleted[id=P]{DE423} \added[id=P]{\citep{Folkner2010}} -} \deleted[id=P]{in the period} is sub-km for barycentric positions of the inner planets \deleted[id=P]{and meet the Gaia DPAC requirements in  positions \deleted[id=P]{with  no  difficulty} \citep{spec2011}}.
For Jupiter, the expected accuracy of \replaced[id=P]{INPOP10e}{the ephemeri\deleted[id=P]{de}s} can not be better than the postfit residuals obtained by comparison to flyby positions which \deleted[id=P]{reach up about} \added[id=P]{are as large as} 2 kilometers. However, in INPOP10e, the Jupiter fitting interval ends in 2001 so \deleted[id=V]{that} no secular trend could have been calibrated during the Gaia period. The situation is similar for Saturn, as its orbit ha\added[id=P]{s}\deleted[id=P]{ve} been highly improve\added[id=P]{d} thanks to the few Cassini observations \replaced[id=A]{. The corresponding}{ and} residuals in geocentric distances \deleted[id=P]{have been} decreased from several hundred kilometers \added[id=A]{for the precedented INPOP versions} to few ten\added[id=P]{s of} meters \added[id=A]{with INPOP10e} but only on the available Cassini period of time \citep{Jones_2014}. For  Uranus  and  Neptune, the DPAC  requirement  \citep{2016NSTIM.104.....F}  of  an  uncertainty  in  positions  of few  kilometers  over  the  Gaia  period was not reached \replaced[id=V]{because of the limited number of observations of these bodies. }{as flyby observations are very limited for theses objects.}

\subsection{INPOP19a} 
\label{inpop19a}

Since \deleted[id=P]{2013} the INPOP10e release \added[id=P]{in 2013},  \deleted[id=P]{new} updated versions of INPOP have been regularly delivered: INPOP13c in 2013 \added[id=A]{with an} improvement of the Mercury orbit \replaced[id=V]{based on}{thanks to}  MESSENGER \deleted[id=P]{radiotracking} \added[id=P]{radio tracking} data \citep{ 2014A&A...561A.115V,INPOP13c}, INPOP15a  \added[id=A]{with new Cassini radio tracking data} \citep{2015CeMDA.123..325F} , INPOP17a \deleted[id=P]{in 2017} \added[id=A]{with an important improvement of the dynamical modeling of the Earth-Moon system \citep{Viswanathan2017}} and finally\added[id=V]{,} INPOP19a \deleted[id=P]{in 2019} \replaced[id=V]{\citep{2019NSTIM.109.....V,2019MNRAS.tmp.3035F}.}{ A detailed description of this version can be found in \cite{2019NSTIM.109.....V} and in \cite{2019MNRAS.tmp.3035F}.}
\added[id=A]{The main differences between INPOP10e and INPOP19a are presented in Table \ref{tab:10ev19a}}

In terms of \deleted[id=P]{data sample} \added[id=P]{dataset}, \replaced[id=V]{INPOP19a}{this version} benefits from \deleted[id=V]{the use of} two major updates: the inclusion of the first nine perijove \added[id=P]{passes} of Juno around Jupiter, and \deleted[id=V]{the} reanalysis of Cassini \replaced[id=V]{radio tracking}{navigation data} around Saturn \citep{2020A&A...640A...7D}\replaced[id=V]{ by}{,} extending the time coverage from 2014 to 2017 (including the final phase of the mission and Titan flyby gravity solutions).

Concerning the dynamical modeling, the main \replaced[id=V]{difference}{change} is coming from the introduction of a set of \deleted[id=P]{3} \added[id=P]{three} \deleted[id=A]{Trans-Neptunian Object (TNO)} rings centered at the \replaced[id=A]{SSB}{\deleted[id=P]{s}\added[id=P]{S}olar system barycenter (SSB)}, located in the ecliptic plane at the distances of 39.4, 44 and 47.5 AU\replaced[id=A]{ and representing the gravitationnal influences of Trans-Neptunian Objects. The total mass of the three ring has been}{, with an} adjusted \replaced[id=A]{to a}{total} mass of 0.061 $\pm$ 0.001 $M_{\oplus}$ \citep{2020A&A...640A...7D,2019NSTIM.109.....V}\replaced[id=V]{. Consequently, Cassini Grand Finale residuals improved significantly.}{, inducing a significant effect on the reduction \added[id=P]{of the} Cassini Grand Finale residuals.} In addition, 343 Main Belt asteroids \deleted[id=P]{has been} \added[id=P]{were} integrated \added[id=P]{and their masses estimated} using a \added[id=A]{Monte Carlo} \deleted[id=P]{Bayesian} procedure described in \cite{2019MNRAS.tmp.3035F}\replaced[id=A]{. This}{ and} leads to an important improvement in the accuracy of the Mars orbit. The well known masses of nine binary TNOs were also included in the \replaced[id=V]{dynamical model}{ephemeri\deleted[id=P]{de}s}. 

\begin{figure}
\begin{center}
\includegraphics[scale=0.2]{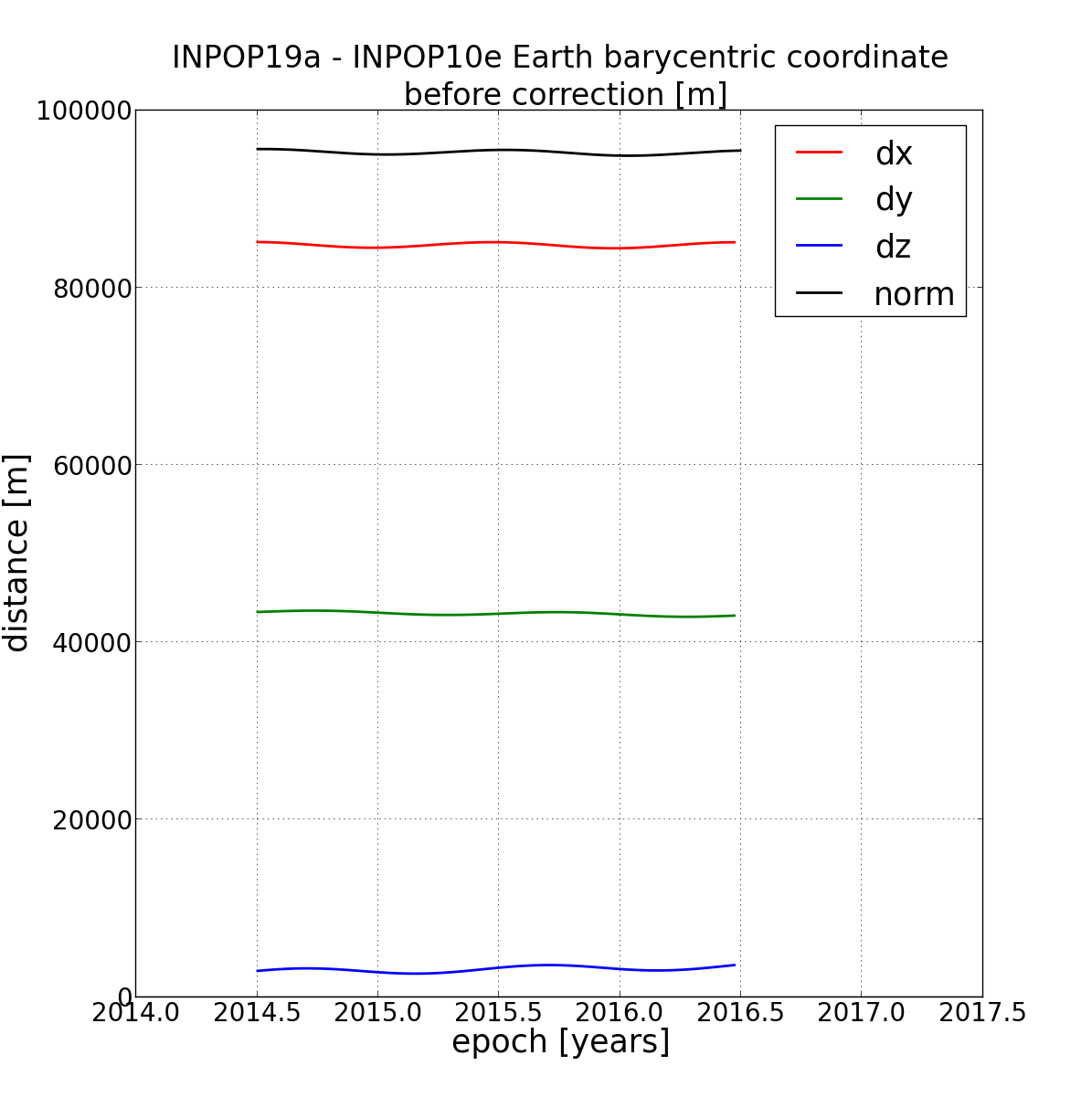}\includegraphics[scale=0.2]{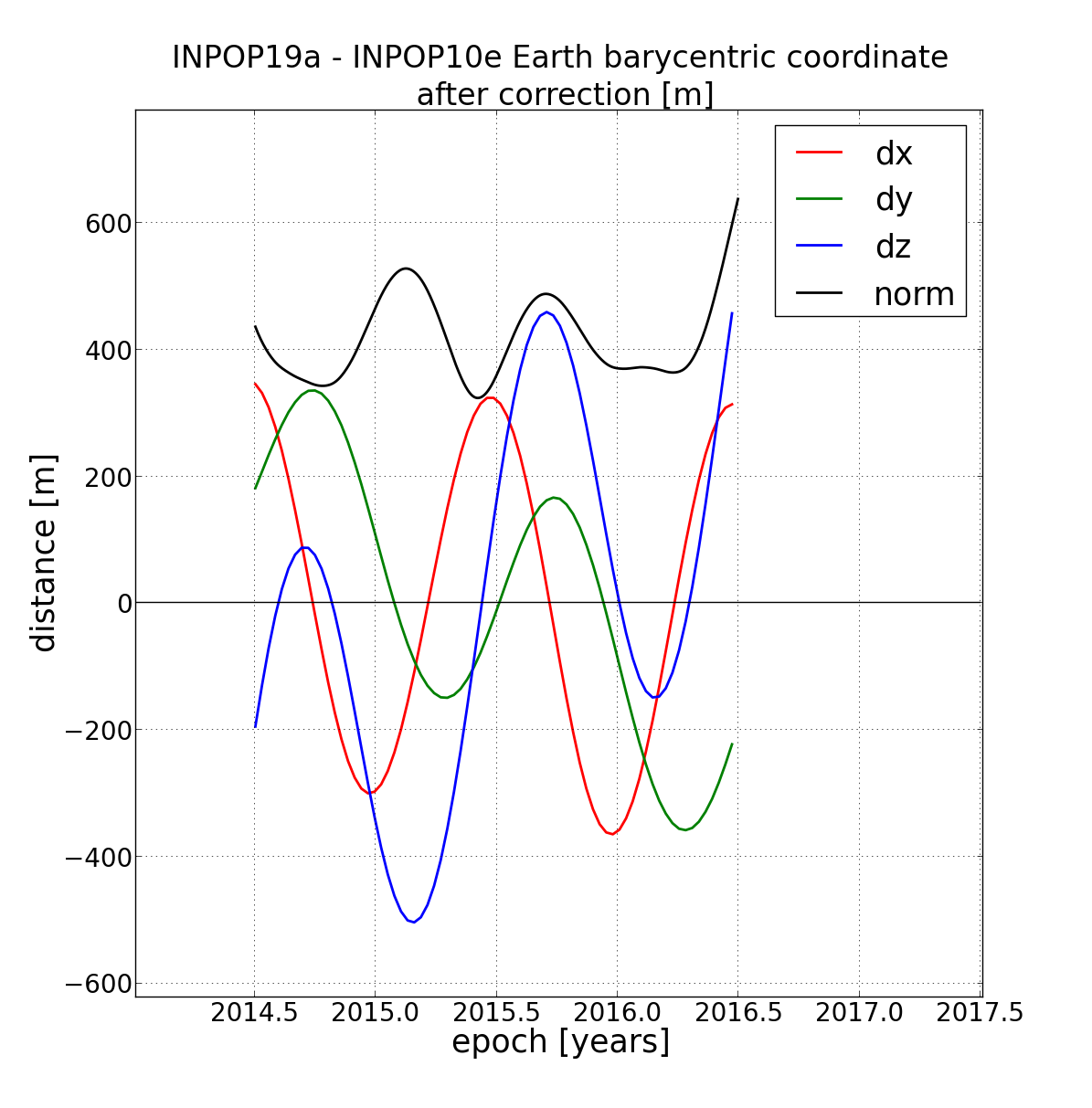}
\end{center}
\caption{SSB-Earth vector difference between INPOP19a and INPOP10e in Cartesian coordinate\added[id=A]{s}. \textit{Left}: before correction by a constant vector. \textit{Right}: After correction by a constant vector based on the mean of the difference. We find dx=84.3 km, dy=44.8 km, dz=3.6 km.}
\label{fig:gaiaomc2}
\end{figure}

Comparing INPOP19a to INPOP10e, we found a difference on the \replaced[id=A]{SSB}{barycenter} position \deleted[id=P]{\added[id=A]{at J2000}} (based on the SSB-Earth vector) of about 95\added[id=A]{~}km \added[id=P]{over the Gaia period} (see Fig. \ref{fig:gaiaomc2}). This \deleted[id=V]{important} difference is mainly due to a pure geometrical effect: the addition of 9 massives TNOs \citep{2019NSTIM.109.....V} between \added[id=A]{INPOP}10e and \added[id=A]{INPOP}19a with planetesimal masses and large \added[id=P]{semimajor} axes (40\deleted[id=P]{ua} \added[id=P]{au}) create\added[id=V]{s} a leverage effect on the \replaced[id=A]{SSB}{barycenter} position. This difference is not constant in time and exhibit\added[id=P]{s} both an oscillation \added[id=P]{(with an amplitude of about 270 meters)} with respect to the Earth orbit period and a regular drift \added[id=A]{(see Fig. \ref{fig:gaiaomc2})}, showing that a dynamical difference \deleted[id=P]{is existing}\added[id=P]{exists} between the two models. As the G\deleted[id=P]{AIA}\added[id=P]{aia} \added[id=A]{satellite} \deleted[id=P]{spacecraft} position\added[id=A]{s} and velocit\added[id=A]{ies} in Gaia-DR2 \added[id=A]{are}\deleted[id=P]{is} given with respect to the \added[id=A]{INPOP10e} SSB \deleted[id=A]{of INPOP10e}, it is clear that \replaced[id=A]{specific cares}{peculiar precautions} must be taken when \replaced[id=A]{substituting INPOP10e with}{using} INPOP19a \added[id=A]{or \added[id=V]{any} other ephemeri\deleted[id=V]{de}s}. Without correction, the 95\added[id=A]{~}km difference implies an error of about \added[id=A]{one} hundred times bigger than the 1 mas accuracy expected by Gaia, and cannot be absorbed during the adjustment of the asteroid initial conditions to the Gaia data.  
\replaced[id=V]{G}{More g}enerally, the position of the \replaced[id=A]{SSB}{Solar system barycenter} depends on the mass\added[id=A]{es} and \deleted[id=V]{\replaced[id=A]{on}{of} the} position\added[id=A]{s} of all the objects \added[id=A]{included in} the ephemerides, implying a strong model dependency of the \added[id=A]{satellite} \deleted[id=P]{spacecraft} position\added[id=A]{s and velocities} provided by \added[id=A]{the} G\deleted[id=P]{AIA}\added[id=P]{aia} release. \replaced[id=V]{We draw attention to the fact that the Gaia team \citep{2018A&Atest} did not apply any correction to the Gaia position when analyzing data with DE431.}{We draw the attention on the fact that in the paper presented by the Gaia team \citep{2018A&Atest}, no correction was applied to the Gaia position\added[id=A]{s} when using DE431 \added[id=A]{planetary ephemeris}.} A quick comparison show\added[id=A]{s} \deleted[id=V]{that} a \replaced[id=V]{discrepancy}{difference} of about 240 meters \replaced[id=V]{in}{on} the \replaced[id=A]{SSB}{Solar system barycenter} position \replaced[id=V]{relative}{given with respect} to the Earth \deleted[id=V]{exists} \added[id=A]{\deleted[id=P]{at J2000} between INPOP10e and DE431}\added[id=P]{ over the Gaia period}. In the worst case, this error \added[id=A]{implies} \deleted[id=P]{could imply} a 0.3 \added[id=A]{mas} \deleted[id=P]{milliarcsecond} difference on the angular position of an object at 1 \added[id=P]{au}\deleted[id=P]{ua} from the \added[id=A]{satellite}\deleted[id=P]{spacecraft}. \replaced[id=V]{Presumably, due}{Due} to the small number of transit\added[id=P]{s} and the short time-span of Gaia-DR2, this error \replaced[id=A]{is}{seems to be} easily absorbed during the fit of the initial condition\added[id=A]{s} of the asteroids - that is why it had remained invisible up to now - but it \replaced[id=V]{would}{will} become critical \replaced[id=A]{with}{in} the future DR3. \added[id=P]{We \replaced[id=V]{believe}{think} that the Gaia \added[id=A]{satellite}\deleted[id=P]{spacecraft} position\added[id=A]{s} would benefit from being provided in the Earth-centered reference frame to avoid this model dependency}. 

\deleted[id=A]{Concerning the accuracy and \added[id=A]{the} residual\added[id=P]{s} of INPOP19a, by \added[id=A]{comparison} \deleted[id=P]{comparing INPOP19a} with the precedent version INPOP17a we obtain a clear improvement of the post-fit residuals especially for Mars, Jupiter and Saturn. For Mars, the \added[id=P]{RMS of the} MRO/MO residual\added[id=P]{s} \deleted[id=P]{improves by 44 percents} \added[id=P]{improves by a factor of 44 \added[id=A]{$\%$} \deleted[id=P]{percents} } on the common interval fit. For Jupiter, the gain reaches \deleted[id=P]{2}\added[id=P]{two} order\added[id=P]{s} of magnitude\deleted[id=P]{s} on the Juno tracking data (from 2km to 20m) while for Saturn, the residual\added[id=P]{s decreased by a factor} \deleted[id=P]{has been divided by} 30 for the Grand finale and by 2.6 for the period between 2006 and 2016. \citep{2019NSTIM.109.....V}. }

In addition to \replaced[id=AF]{the}{this} planetary adjustment, we \deleted[id=P]{decide to add}add\added[id=P]{ed} to INPOP19a the positions and velocities of the 14099 asteroids observed by G\deleted[id=P]{AIA}\added[id=P]{aia} \added[id=A]{and provided by Gaia-DR2}. 
\replaced[id=AF]{The asteroid orbits are integrated with the same code and integrator as INPOP19a using the same initial conditions but with}{Starting from the INPOP19a planetary solution, the \added[id=A]{asteroid} \deleted[id=P]{SSO} orbit determination \deleted[id=P]{has been}\added[id=P]{was} performed independently, in} a Newtonian formalism, including the perturbations of the Sun and main planets but with a reduced number of perturbing asteroids \added[id=A]{compared to INPOP19a}. \added[id=A]{In terms of perturbing asteroids,} for a sake of comparison, we chose the same list of \replaced[id=A]{16 perturbers}{perturbing asteroids (16)} as in \cite{2018A&Atest}. \added[id=P]{The masses of the 16 perturbers included in the dynamical model are presented in table \ref{tablemass} and were extracted from \cite{2019MNRAS.tmp.3035F}} \deleted[id=P]{ \replaced[id=A]{We also tested}{However, after testing different} alternative lists of perturbers, \replaced[id=A]{but}{and} due to the limited interval of time covered by the G\deleted[id=P]{AIA}\added[id=P]{aia} data, no differences are noticed on the \added[id=A]{Gaia} residuals after the fit.}

\begin{table}[h!]
\centering
\caption{List of the 16 perturbing asteroids used in \cite{2018A&Atest} with their corresponding mass in INPOP19a. Stars indicate the five objects that were also observed by Gaia and available in DR2.}
\begin{tabular}{||c c||} 
\hline
Asteroid number & GM \\
- & [$10^{18} AU^3.d^{-2}$] \\ [0.5ex] 
 \hline\hline
1 & 139643.532 \\
2 & 32613.272 \\
3 & 3806.229 \\
4 &  38547.977 \\
6 & 986.372 \\
7 &  1833.933 \\
10 & 11954.671 \\
15 &  3936.840 \\
16 & 3088.668 \\
29 & 2103.023 \\
52* & 2308.950 \\
65 & 2990.906 \\
87* & 2726.311 \\
88* & 0.055 \\
511* &  6637.430 \\
704* & 4737.367 \\[1ex] 
 \hline
\end{tabular}
\label{tablemass}
\end{table}

Finally, we verify that the asteroid orbital improvements \replaced[id=V]{resulting from}{brought by} the Gaia observations on the massive \added[id=A]{asteroids} \deleted[id=P]{SSOs} \added[id=A]{included in the INPOP19a modeling} do not have any significant impact on the previously obtained planetary orbits\replaced[id=V]{, as such}{ such as} no iteration \replaced[id=A]{is}{was} \replaced[id=V]{needed}{necessary} between \deleted[id=V]{the} asteroid \added[id=V]{and planetary} adjustment procedure\deleted[id=V]{and the planetary one}. 

{\obf{We also check that no residual rotation has been observed between the planetary ephemerides obtained with and without the Gaia observations. On Table \ref{angles}, one can find Euler angles fitted over the two ephemerides for different cases, depending which planetary orbits are considered. If all the orbits including the outer planet ones are considered, the mis-alignement of the INPOP reference frame axis (by definition, the ICRF , without considering the VLBI observation uncertainties) with the Gaia reference frame (GRF) is not statistically significant with {\obf{standard deviation}} smaller than 1 mas. However if we consider only the orbits fitted over very accurate observations such as the inner planets, Jupiter and Saturn, then the Euler angles turn out to have significant values but at the level of few $\mu$as.  This is far below the uncertainty of the alignement between DR2 GRF and ICRF3 of about 20 to 30 $\mu$as as obtained by (\cite{refId0}).  
We can then conclude to a good alignement of the INPOP reference axis relative to the DR2 GRF.}}

\begin{table}
\caption{Euler angles fitted by comparing two planet ephemerides only different by the asteroid orbits used for computing their perturbations on the planet orbits: one being fitted over the Gaia DR2 and one obtained from the astorb data base.}
\begin{tabular}{l c c c}
\hline
& $\theta$ & $\psi$ & $\phi$ \\
& $\mu$as & $\mu$as & $\mu$as \\
All planets & $-98 \pm 1508 $ & $1.0 \pm  45 $& $253 \pm 3971$\\
Inner planets & $ 1.16 \pm 0.20 $& $-0.08 \pm 0.62$& $-1.50 \pm 0.150$\\
Inner planets + Jupiter & $1.23 \pm 0.16 $& $-1.0 \pm 0.22$& $-0.128 \pm 0.69$\\
Inner planets + Jupiter + Saturn  & $1.83 \pm 0.80$&$1.057 \pm 0.053$ &$-0.23 \pm 2.56$ \\
Outer planets & $-55\pm 629$ & $-30 \pm 495$& $373 \pm 5051$\\
\hline
\end{tabular}
\label{angles}
\end{table}


\section{\replaced[id=P]{\deleted[id=P]{Correcting} Asteroid orbits \added[id=A]{with}\deleted[id=P]{using} Gaia-DR2} {Adjustment of asteroid orbit using Gaia DR2}}

\label{adjustment}

\subsection{Method}

\label{fit}

\deleted[id=P]{14099 orbits have  been integrated}\added[id=P]{We integrated 14099 orbits} with INPOP \replaced[id=V]{using}{following} the procedure described \added[id=A]{previously and computed observables as given }in \cite{2018A&Atest}. 
\added[id=A]{Concerning the positions of the Gaia \added[id=A]{satellite}, as the SSB is shifted between INPOP10e and INPO19a (see Sect. \ref{inpop19a}), we \added[id=A]{apply a preliminary transformation from Gaia barycentric positions to Gaia geocentric ones} \deleted[id=P]{made a preliminary conversion of the Gaia spacecraft position to that relative to the geocenter in INPOP10e} before switching to the new ephemeris. The idea is to consider the Earth-Gaia vector provided with the \added[id=A]{DR2} Solar system modeling \deleted[id=P]{used to make Gaia-DR2} (INPOP10e) as an invariant, independent \added[id=A]{from} \deleted[id=P]{of} the model \added[id=A]{. This is} \deleted[id=P]{, which is} justified \added[id=A]{because the satellite} \deleted[id=P]{by the fact than spacecraft} positions are usually based on ground-based telemetry. Therefore, for each Gaia observation, we simply report the Gaia geocentric position obtained with INPOP10e to INPOP19a, and \added[id=A]{we then avoid} \deleted[id=P]{thus avoiding} the model dependency intrinsically present in the barycentric coordinates.}
\added[id=A]{At this step, after a first comparison between computed observables and Gaia observations, an optimisation of the asteroid orbital initial conditions is required.}
\deleted[id=P]{Therefore,} The goal \deleted[id=V]{of the least square\added[id=A]{s}} is to find the optimal \added[id=A]{set of} \deleted[id=P]{asteroids} initial conditions \added[id=P]{for the considered asteroids} by minimizing the sum S of weighted squared residual\added[id=A]{s} \added[id=V]{using least square technique} \added[id=P]{:} 

\begin{equation}
S= \xi^T W \xi
\label{S}
\end{equation}

\noindent\added[id=P]{w}here \deleted[id=P]{m is the number of observations}$\xi$ is the vector of the residuals \deleted[id=P]{(Observation minus simulations)} \added[id=P]{(Observed - Computed \added[id=A]{)}}\added[id=V]{,} and W is the weight matrix.

As the observations were acquired in the AL/AC Gaia \deleted[id=P]{proper} frame and delivered in the right ascension ($\alpha$) and declination ($\delta$) coordinates \replaced[id=A]{given}{expressed} in the barycentric reference system (BCRS), \added[id=V]{a} strong correlation\added[id=A]{s} exist\deleted[id=P]{s} between the random error\added[id=A]{s} in \added[id=A]{$\alpha$}\deleted[id=P]{ra} and \added[id=A]{$\delta$}\deleted[id=P]{dec}\replaced[id=A]{. These have to be}{ that must be} fully \replaced[id=A]{accounted for}{taken into account} during the orbit determination process.

\deleted[id=P]{computed as the invert \added[id=P]{i.e., the inverse} of the random covariance matrix:}

The general solution is \deleted[id=A]{then} \replaced[id=A]{obtained}{given} using a Gauss-Newton iterative procedure where each step requires the resolution of the well-known normal equation: 

\begin{equation}\label{gn}
\Delta x=-(J^TWJ)^{-1}J^TW\xi
\end{equation}

\noindent \added[id=V]{where} $\Delta$x is the vector of the differential correction\added[id=A]{s} that must be applied to the initial condition\added[id=A]{s} from one iteration to the next. $J$ \replaced[id=V]{is}{-} the Jacobian matrix \replaced[id=V]{containing}{of our model - is} the first derivatives of the residual\added[id=A]{s} with respect to the \deleted[id=P]{adjusted}\added[id=P]{estimated} parameters:

\begin{equation}
J_{ij}=\frac{\partial\xi_i}{\partial x_i}
\label{newton}
\end{equation}

\noindent with $i\le 2\times m$ \added[id=P]{(m is the number of \added[id=A]{$(\alpha,\delta)$} \deleted[id=P]{(ra,dec)} observations)} and $j\le 6\times n$ (n is the number of adjusted \replaced[id=A]{orbits}{asteroids}). In practice, the derivatives are computed numerically\deleted[id=P]{by making the difference between two simulations starting from the same initial condition shifted respectively by +dp and -dp (typical value for |dp| is 10e-7 au for positional coordinates and 10e-6 au/years for velocities)}. 
\replaced[id=O]{We obtain an interesting shape for the Jacobian matrix by sorting both the residual vectors and the adjusted paramaters with the same asteroid order starting from the $p$ perturbing asteroids observed by Gaia}{An interesting shape for the Jacobian matrix can be obtained by sorting both the adjusted parameters and the residuals with the same asteroid order starting from the \replaced[id=A]{$p$ asteroids observed by Gaia and perturbing all the other considered objects}{\deleted[id=P]{5}\added[id=P]{five}  perturbing bodies observed by Gaia}}. In this case, the first $6\times p$ columns of the matrix depict\deleted[id=P]{s} the influence of the perturbing bodies on \replaced[id=P]{the whole residuals}{\added[id=A]{the} other asteroid residuals}, whereas the other part of the matrix is a diagonal block wise submatrix, where each block depicts the influence of the initial conditions of a perturbed asteroid on \deleted[id=P]{his}\added[id=P]{its} own residuals. If we call $J_i$ \replaced[id=A]{this latest block}{this kind of block}, and $P_i$ the derivative\added[id=A]{s} of the residual\added[id=A]{s} of the $i$th asteroid with respect to the initial coordinates of the \replaced[id=A]{$p$}{\added[id=A]{5}} perturbers\added[id=V]{,} we obtain:

\begin{equation}\label{fullmatrix}
J
=
\begin{bmatrix}
\boxed{J_1} & & & & & \\
\boxed{P_2} & \boxed{J_2} & & & & \\
\boxed{P_3} & & \boxed{J_3} & & 0 & \\
\boxed{P_4} & & & \boxed{J_4} & & \\
\vdots & & 0 & & \ddots & \\
\boxed{P_n} & & & & & \boxed{J_n} \\
\end{bmatrix}
\end{equation}

The same formalism is applied to the residual vector $\xi$ and the weight matrix W:

\begin{equation}
\xi =\begin{pmatrix}\xi_1\\\xi_2\\...\\\xi_i\\\xi_n\end{pmatrix}
\end{equation}

\begin{equation}
W
=
\begin{bmatrix}
W_1 &  &  &
&  \\
 & W_2 &  & 0
&  \\
 &  & \ddots
& & \\
& 0 & 
& W_i &  \\
&  & 
&  & W_n
\end{bmatrix}
\end{equation}

\noindent\added[id=P]{,w}here $\xi_1$ and $W_1$ are respectively the residual vector and \deleted[id=V]{the} weight matrix for the  \replaced[id=A]{$p$}{\added[id=A]{5}} perturbers and $\xi_i$, $W_i$ for the $i$th pertubed body (starting at i=2).
For inverting such a system, a strategy of block-wise inversion using the Schur complement \replaced[id=P]{\citep{zhang05}}{(\added[id=A]{REFERENCE ?})} was performed,  and a global solution for the normal equation\added[id=A]{s} is given by: \\

\begin{equation}
\label{eq1}
   \left \{
   \begin{array}{r c l}
      \Delta x_1  & = & -COV(J_1^TW_1\xi_1 + B) \\
      \\
      \Delta x_i  & = & -(J_i^TW_iJ_i)^{-1}J_i^TW_i(\xi_i-P_ix_1)\,\,\,\,\, for\, (i\geq2)
   \end{array}
   \right .
\end{equation}

\noindent\added[id=P]{,w}ith: 

\begin{equation}
COV=(J_1^TW_1J_1+A)^{-1}
\end{equation}

\begin{equation}
\label{eq3}
\begin{array}{cc} A=\sum_{i=1}^{n} C_i P_i\,\, & \,\,B=\sum_{i=1}^{n} C_i \xi_i\\ \end{array}
\end{equation}

\noindent and

\begin{equation}
C_i=P_i^T W_i [1-J_i(J_i^T W_i J_i)^{-1} J_i^T W_i]
\end{equation}

\deleted[id=P]{in prediction of the huge}\added[id=P]{In \added[id=A]{the context }\deleted[id=P]{light} of the expected} amount of data that will be release\added[id=A]{d} \added[id=A]{with the} \deleted[id=P]{in} Gaia DR3, the solution given in Eq. \ref{eq1} is convenient because it only handles small matrices that can be computed in an iterative way (see Eq. \ref{eq3}). We note that if all \deleted[id=V]{the} $P_i$ are set to zero (no perturbers), Eq. \ref{eq1} is \added[id=V]{then} equivalent to $n$ independent normal equations where each asteroid is adjusted without taking into account the others. 

In practice, the $P_i$ perturbing terms (caused by the small variation\added[id=A]{s} of the perturber \replaced[id=A]{orbital initial conditions}{positions}) are very small and the solution provided by \replaced[id=A]{a system where all the asteroids are independent}{such a system} would be very close to the one \added[id=A]{found} \deleted[id=P]{find} with this rigorous method \added[id=A]{accounting for the perturbations of main perturbers}. This could be easily seen by looking at the second equation of \ref{eq1}, where the solution is the classical one applied to a residual corrected from the perturbation induced by the perturbers.

\added[id=P]{A quick estimation of the impact of the position \added[id=A]{uncertainty} of a perturber on a perturbed \added[id=A]{orbit} could be computed by \replaced[id=V]{estimating}{considering} the derivative of the acceleration with respect to the position. A \added[id=A]{perturber} position error $dr$ \deleted[id=P]{in the perturber} causes an acceleration error proportional to $dr/r$, where $r$ is the distance between the perturber and the perturbed asteroid. For a 100 km error at 1 au the relative effect \added[id=A]{$\delta a /a$} is \replaced[id=A]{smaller than}{<} $10^{-6}$. \replaced[id=V]{As a comparison, the error in the mass of the perturber (in this case, a few percent) is likely to be much larger, so the effect of \added[id=A]{the perturber} trajectory errors on \replaced[id=A]{perturbed}{perturber} residuals is unlikely to be \replaced[id=A]{significant}{meaningful} unless the mass is estimated as well.}{As a comparison, the \added[id=A]{relative uncertainty on} \deleted[id=P]{error in} the mass of the perturber is likely \deleted[id=P]{far} greater \deleted[id=P]{than that} (few percent\added[id=A]{s}) so that in our case,  there is little value in considering the effect of perturber trajectory errors on \added[id=A]{perturbed} \deleted[id=P]{perturber} asteroid residuals unless the mass is estimated as well.} However, as this formalism is adapted for the adjustment of any global parameter (it requires only to add few columns in the matrix $P_i$), it will be particularly convenient in the perspective of Gaia DR3, where the mass\added[id=A]{es} of the biggest objects would be probably \added[id=A]{measurable} \deleted[id=P]{ detectable} thought their \added[id=A]{perturbations} \deleted[id=P]{impact} on the trajector\added[id=A]{ies} of multiple asteroids \citep{2008P&SS...56.1819M}}   
\deleted[id=P]{However, as This formalism is also particularly adapted for the adjustment of any global parameter, like the masses of the biggest objects which would be probably attainable with Gaia DR3: it requires only to add few columns in the matrix $P_i$. This convenient feature was applied in our case in order to correct the barycentric position of the Gaia spacecraft expressed in the INPOP19a frame (see Sect. \ref{inpop19a}). It consists in adding the three coordinates of a translation vector related to the G\deleted[id=P]{AIA}\added[id=P]{aia} position in the global adjustment of the initial condition. The idea is to constraint the spacecraft position using the minimization of the residual as the criteria. We find dx=86.6 km dy=49.3 km dz=5 km that must be compared to the mean difference on the Earth-barycenter position obtained by a direct comparison between INPOP10e and INPOP19a before the fit dx=84.3 km, dy=44.8 km, dz=3.6 km (see Sect. \ref{fig:gaiaomc2}). It is explained by the fact that the correction tends to maintain the  Earth-Spacecraft vector as an invariant between INPOP10e (where the data were processed) and INPOP19a (where the adjustment was processed). The difference of 5.2 km between the two can be explained by the fact than during the fit, the adjustment of the initial conditions mechanically implies a shift of the barycenter  (especially, 2 TNOs were adjusted on Gaia observation) so that the correctional vector must move as well. Moreover, as the adjustment of the corrective vector is based on the Gaia observations, it directly depends both on the quality and the quantity of the observations, and on its correlation with the fitted CI (for the same angular accuracy, an error on the spacecraft position will have a greater impact on Near-Earth asteroid residuals, than on Trans-Neptunian ones). Using both of the two corrective vectors, we find no significant difference in the global post-fit residuals which tend to prove that, using only the Gaia-DR2 observations, a 5km uncertainty on the spacecraft position can still be globally absorbed by the adjustment of the asteroid initial conditions. Finally, even after the correction by a constant vector relative to the Solar system barycenter, Figure \ref{fig:gaiaomc2} (right panel) shows that a dynamical difference between the two model is still existing (of about 417m) so that the Earth-spacecraft position cannot be rigorously maintained constant between INPOP10e and INPOP19a by only using a constant translation of the spacecraft position. This residual error - easily absorbed so far - must be absorbed in the future by a time-dependent additive correction.}

\deleted[id=A]{\added[id=P]{Concerning the  the Gaia \added[id=A]{satellite}\deleted[id=P]{spacecraft} position\added[id=A]{s}, as the \added[id=A]{Solar system} barycenter is shifted between INPOP10e and INPO19a (see Sect. \ref{inpop19a}), we \added[id=A]{apply a preliminary transformation from Gaia barycentric positions to Gaia geocentric ones} \deleted[id=P]{made a preliminary conversion of the Gaia spacecraft position to that relative to the geocenter in INPOP10e} before switching to the new ephemeris. The idea is to consider the Earth-Gaia vector provided with the \added[id=A]{DR2} Solar system model \deleted[id=P]{used to make Gaia-DR2} (INPOP10e) as an invariant, independent \added[id=A]{from} \deleted[id=P]{of} the model \added[id=A]{. This is} \deleted[id=P]{, which is} justified \added[id=A]{because the satellite} \deleted[id=P]{by the fact than spacecraft} positions are usually based on ground-based telemetry. Therefore, for each Gaia observation, we simply report the Gaia geocentric position obtained with INPOP10e to INPOP19a, and \added[id=A]{we then avoid} \deleted[id=P]{thus avoiding} the model dependency intrinsically present in the barycentric coordinates.}}

\subsection{Results of the Gaia fit}

\label{resultgaia}

\begin{figure}
\begin{center}
\includegraphics[scale=0.19]{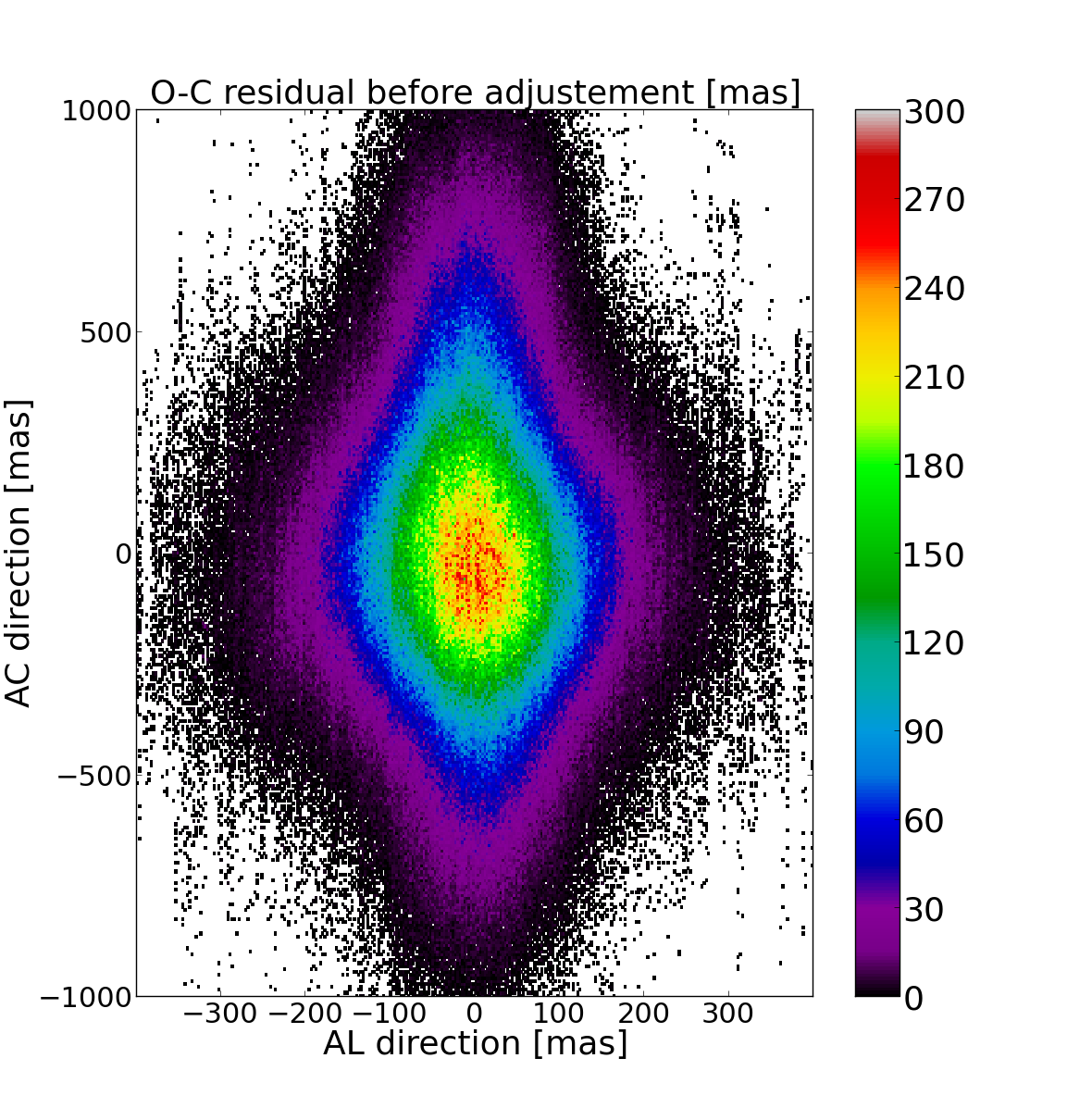}\includegraphics[scale=0.18]{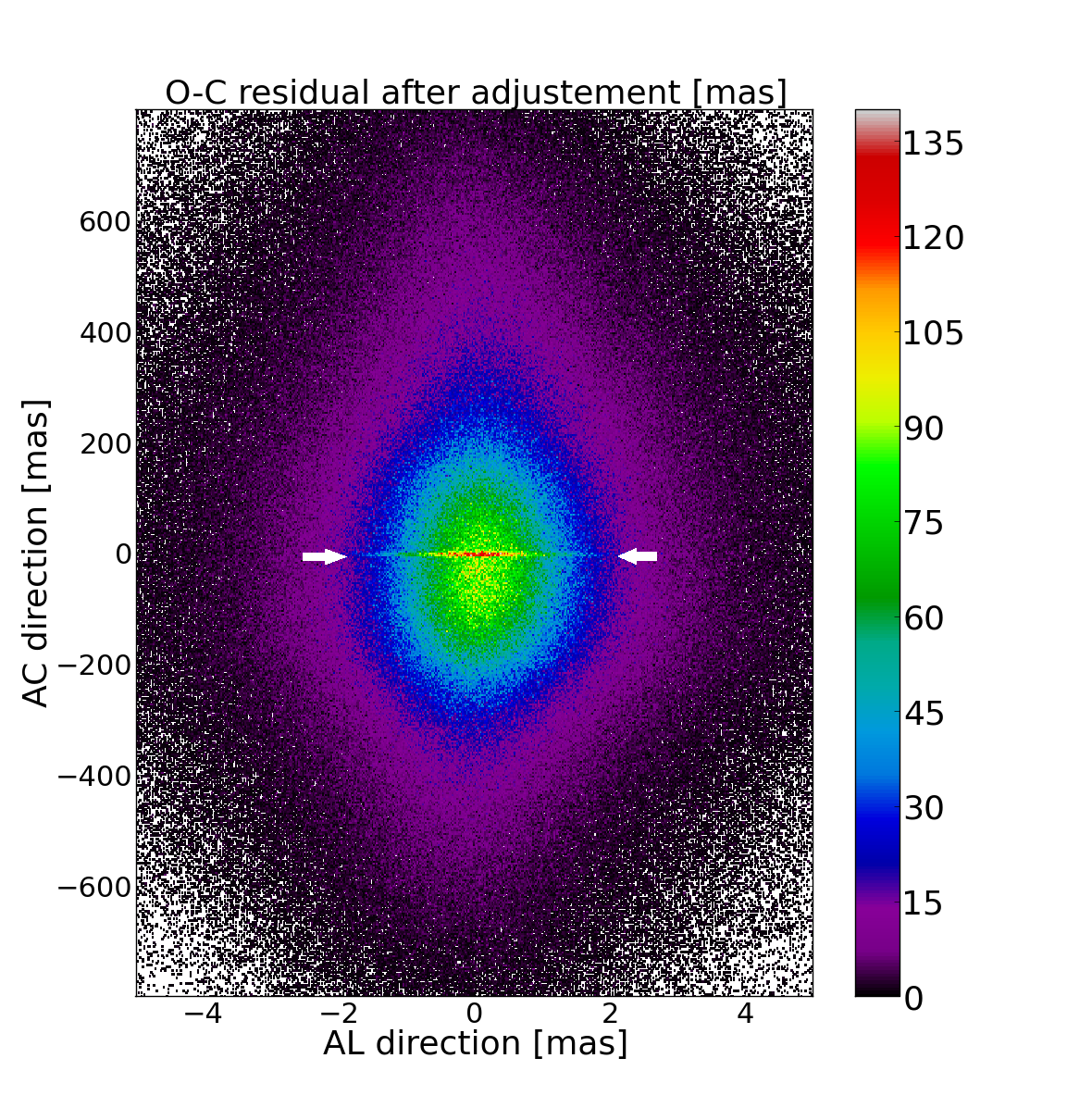}
\end{center}
\caption{Density plot of the residuals in the (AL,AC) plane expressed in milliarcsecond\added[id=P]{. Left panel: before \deleted[id=P]{and after} the adjustment of the initial conditions. Right pane: after the adjustment of the initial conditions.} The colorbar and the axis range were chosen to be directly comparable with Fig.19 of \cite{2018A&Atest}. \added[id=P]{The white arrows point the over-density created by residuals of asteroids with G<13}. }
\label{gaiaomc3}
\end{figure}

\begin{figure}
\begin{center}
\includegraphics[scale=0.25]{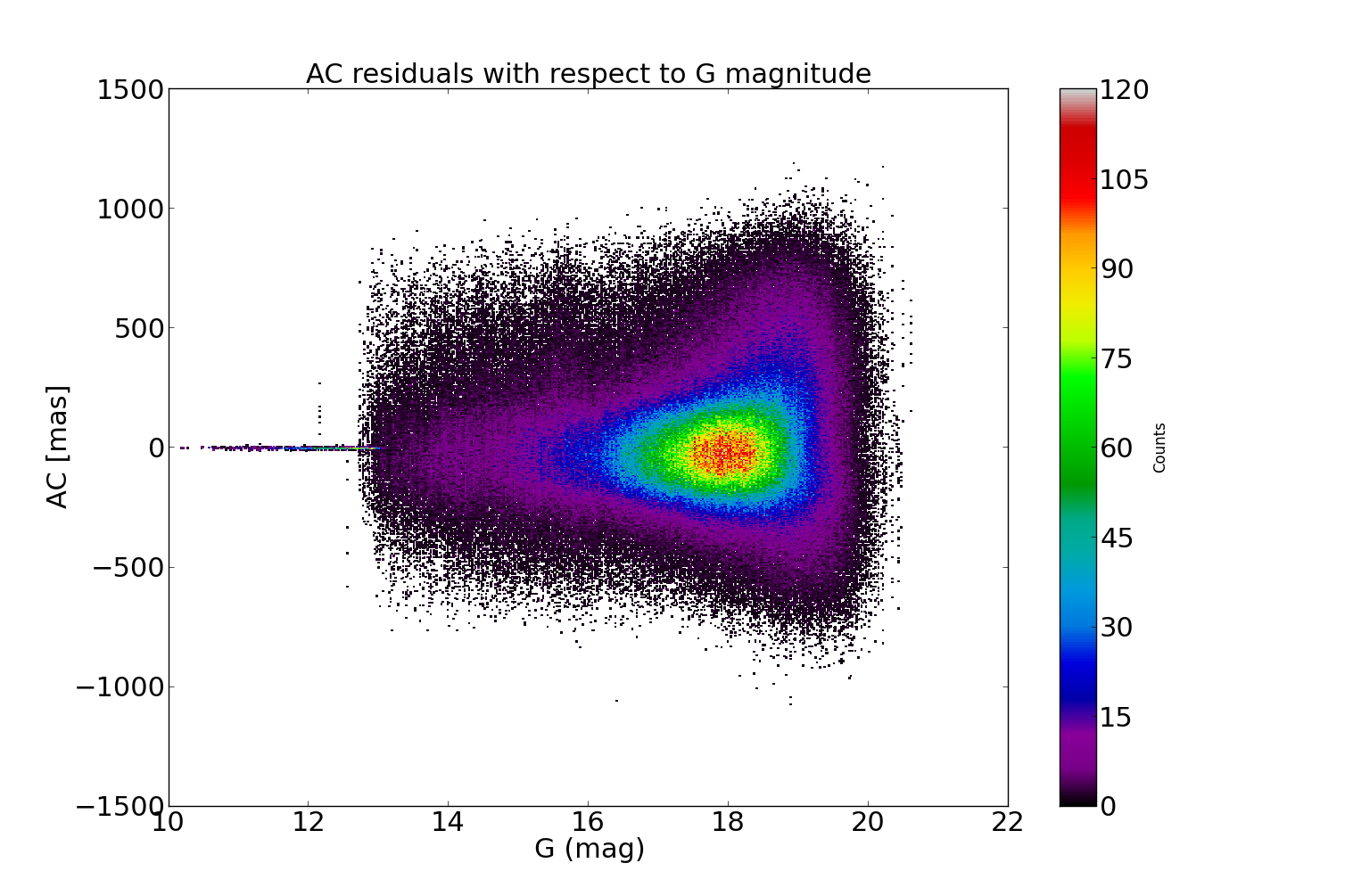}\includegraphics[scale=0.25]{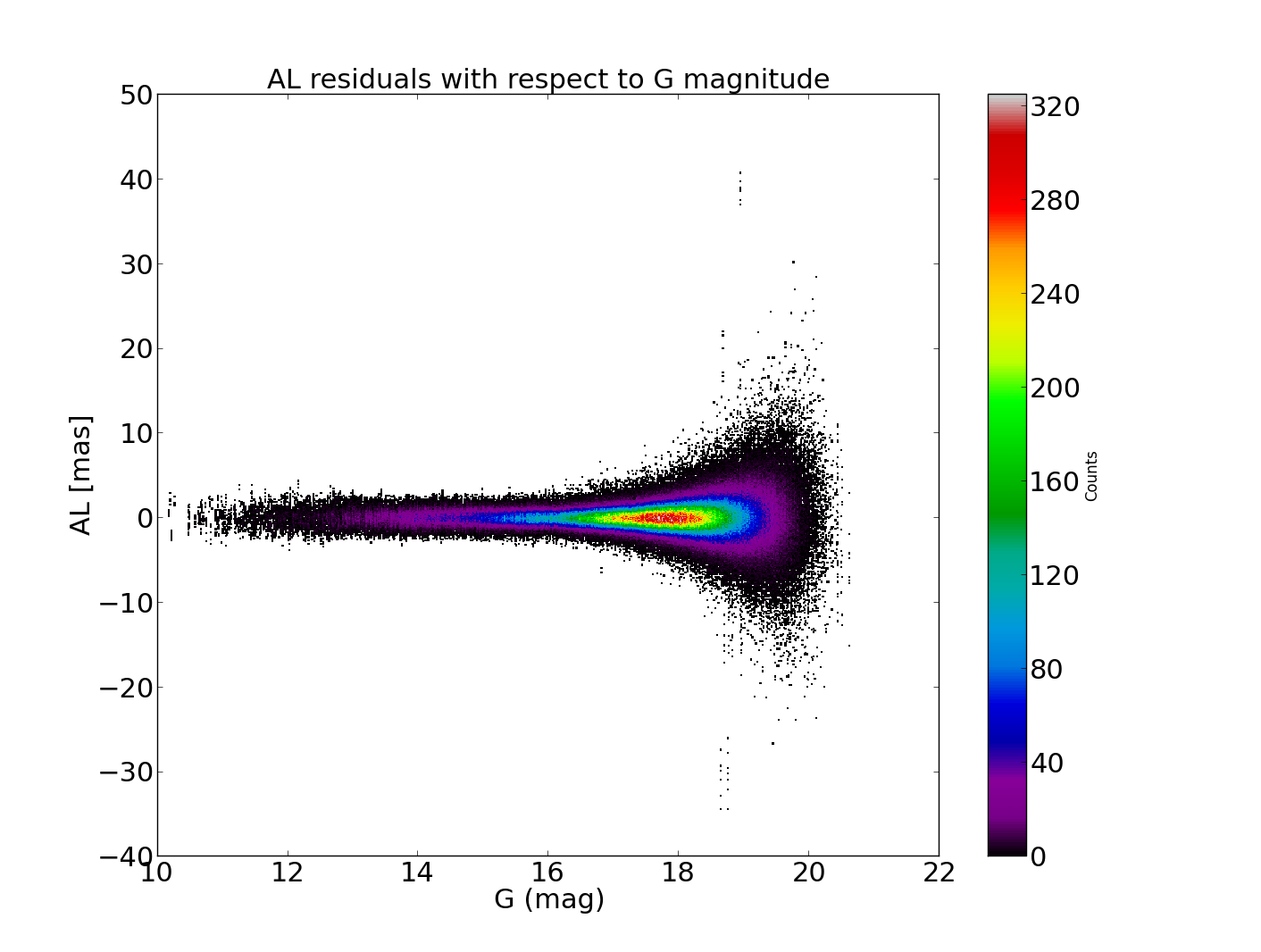}
\end{center}
\caption{Density plot of the residuals in AL and AC with respect to the magnitude G after the adjustment of the initial conditions. }
\label{gaiaomc4}
\end{figure}

Fig. \ref{gaiaomc3} presents the residuals obtained before and after the fit of the 14099 asteroids using INPOP19a. \deleted[id=P]{it}\added[id=P]{The fitted dataset} includes 1 977 702 observations \replaced[id=V]{from}{corresponding to} 287940 transits. 98\% of the AC residuals fall in the interval \added[id=A]{$\pm$ 800 mas} \deleted[id=P]{[800 800] milliarcsec}. 96\% of the AL residuals fall in the interval \added[id=A]{$\pm$ 5 mas} \deleted[id=P]{[-5,5] milliarcsec} and 53\% are at \added[id=V]{the} sub-milliarcsec\deleted[id=V]{ond} level.  A comparison of these results with those \replaced[id=A]{published}{presented} in  \cite{2018A&Atest} is presented in Table \ref{tablecompare}\replaced[id=V]{. Both results are highly consistent}{, confirming the \deleted[id=P]{goodness of the fit} \added[id=P]{agreement between the two}} \added[id=A]{, at least in the AL direction}.  The peak around 0, visible in \added[id=A]{the right-hand side figure of} Fig. \ref{gaiaomc3} for the AC direction, is related to the magnitude dependenc\replaced[id=A]{y}{e} of the residual\added[id=P]{s} and \replaced[id=A]{gathers}{is composed \added[id=A]{by} \deleted[id=P]{of}} objects brighter than G=13 for which the accuracy in the AC direction reaches \added[id=A]{the} milliarcsecond level. This \added[id=P]{behavior} \replaced[id=V]{is also apparent in}{can also be seen on} Fig \ref{gaiaomc4}\added[id=V]{,} showing the residual distribution as a function of the G magnitude.

\begin{figure}
\begin{center}
\includegraphics[scale=0.25]{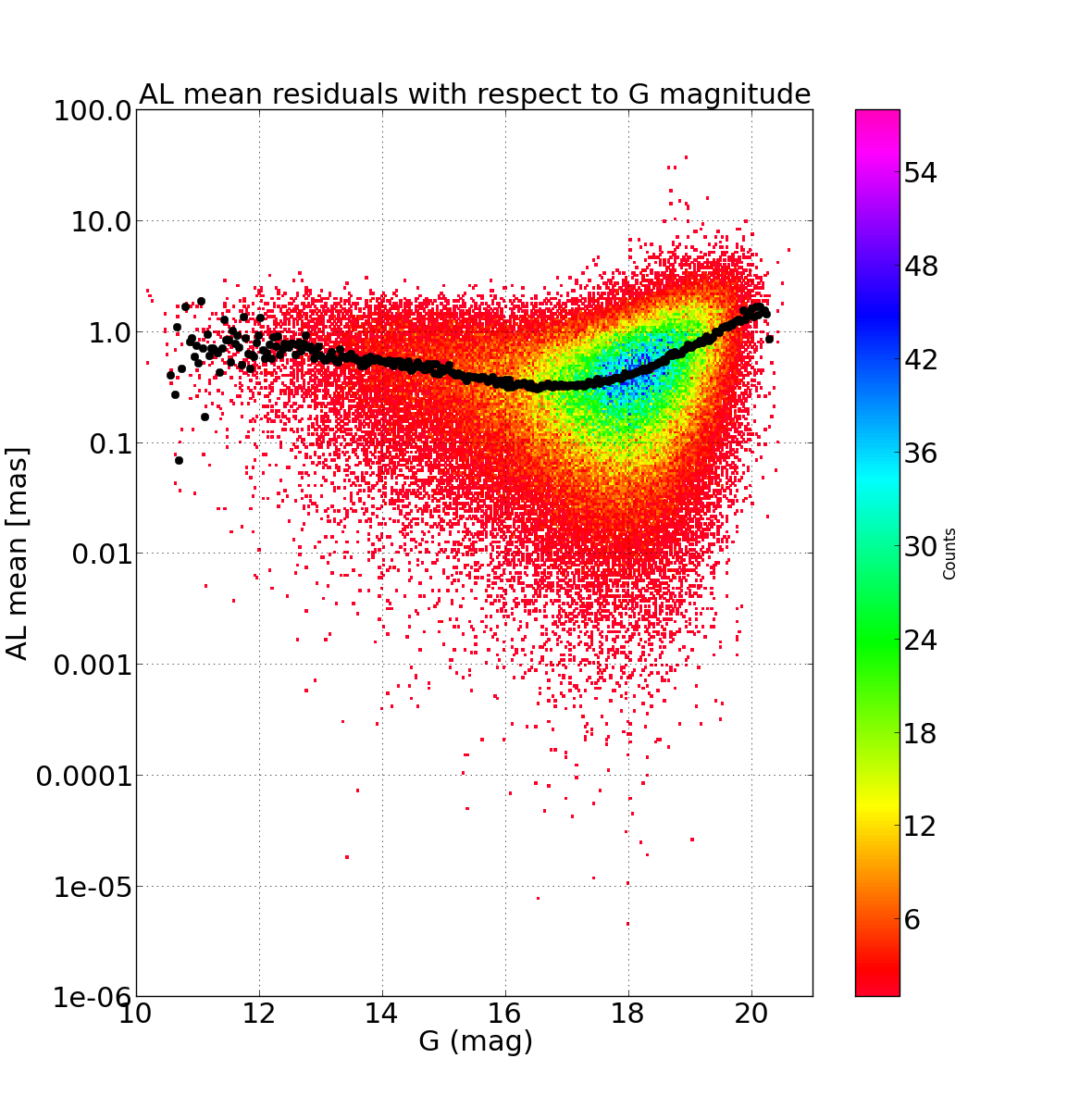}\includegraphics[scale=0.25]{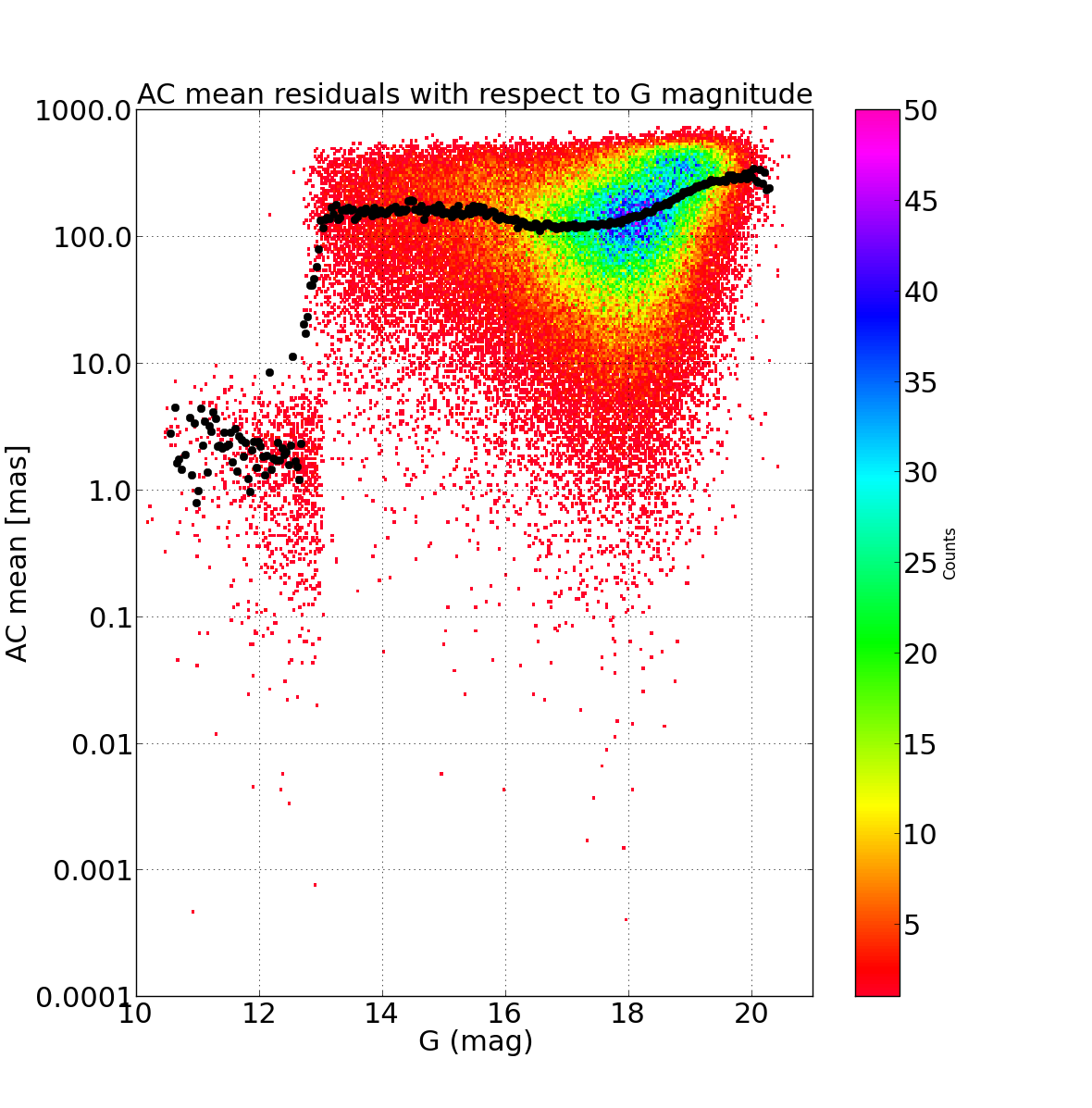}\\
\includegraphics[scale=0.25]{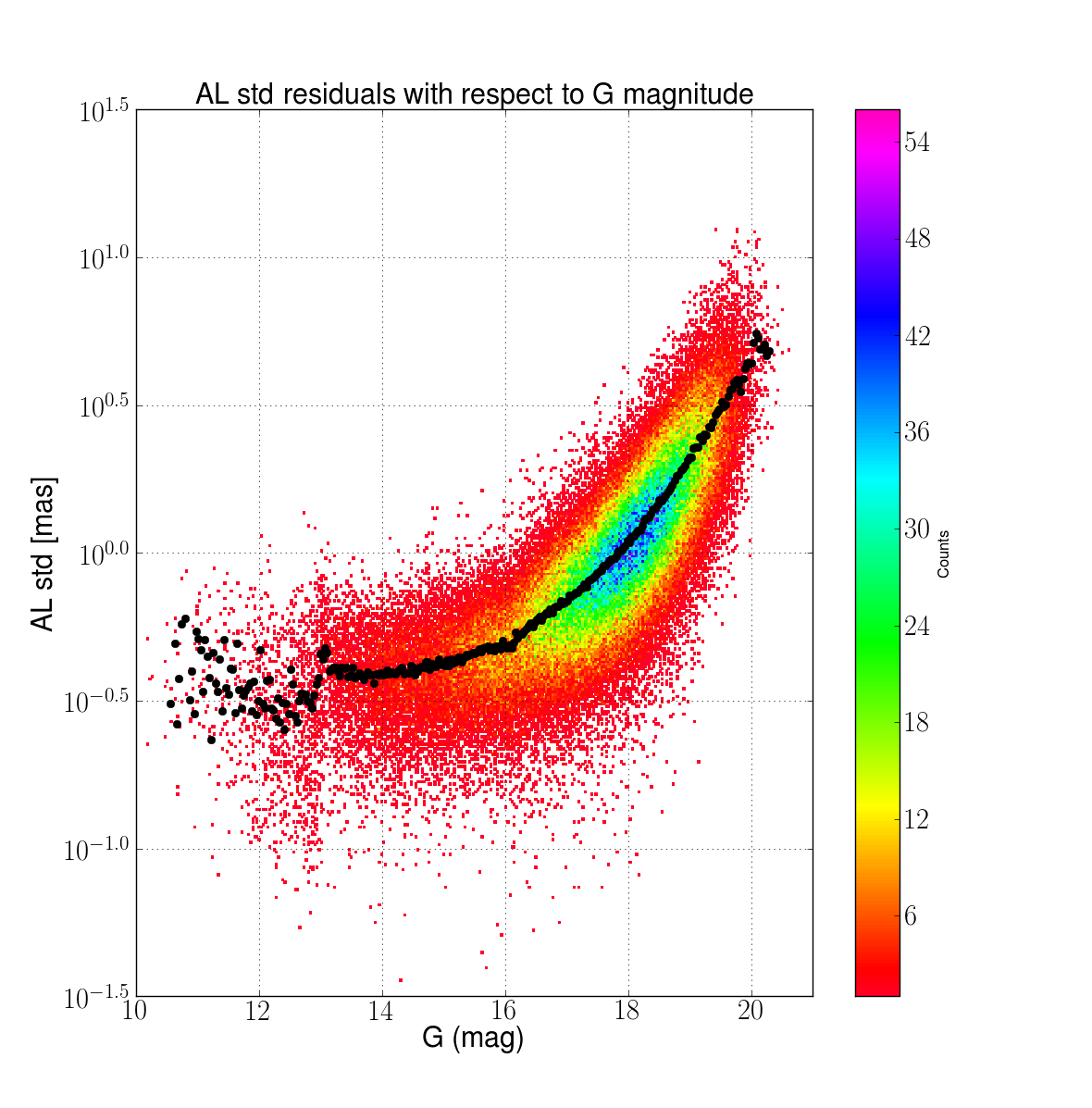}\includegraphics[scale=0.25]{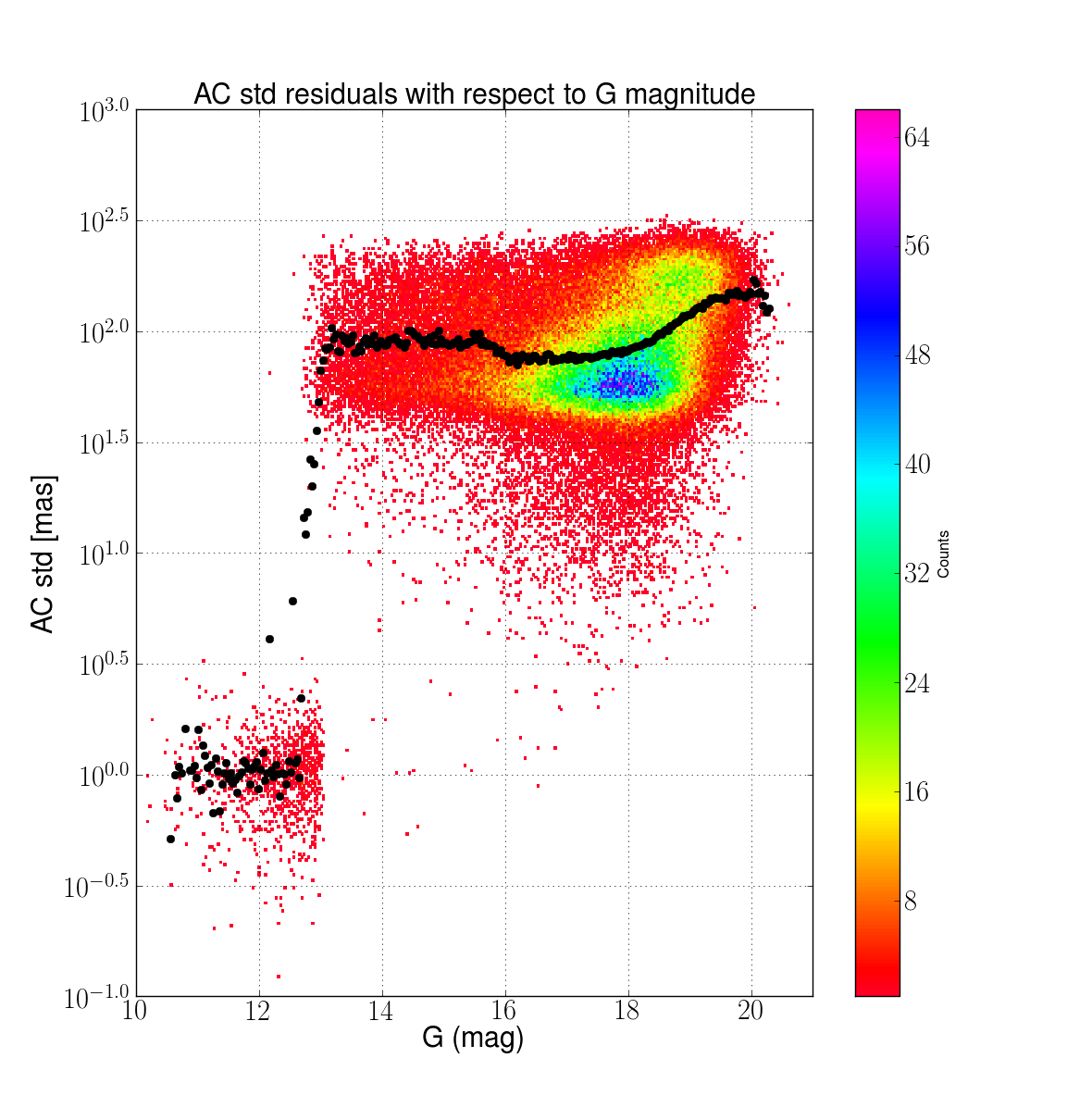}
\end{center}
\caption{Density plots of the mean residuals (first row) per transit in AL and AC with respect to the magnitude G after the adjustment of the initial conditions and density plots of the standard deviation of the residuals (second row) per transit in AL and AC with respect to the magnitude G after the adjustment of the initial conditions.}
\label{transit1}
\end{figure}

\added[id=P]{Similarly to \added[id=A]{Figs.} \deleted[id=P]{figure} 27 and 28 in \cite{2018A&Atest}}, \deleted[id=V]{we show on} Fig. \ref{transit1} \deleted[id=A]{and \added[id=A]{Fig.} \ref{transit2}} \added[id=V]{shows} the absolute value\added[id=A]{s} of the mean and standard deviation\added[id=A]{s} per transit. \replaced[id=A]{These representations are}{This approach is} more convenient than the one used on Fig. \ref{gaiaomc3}, as \replaced[id=A]{they put}{it \deleted[id=P]{is supposed to} put\added[id=A]{s}} into evidence the existence of systematic errors at a transit level (classically, the mean\added[id=A]{s} of the residual\added[id=A]{s} per transit) and provide a better estimation of the debiased residuals (classically, the standard deviation\added[id=A]{s} of the residual\added[id=A]{s} per transit),\added[id=A]{to} \deleted[id=P]{to which must} be compared \added[id=A]{with} the expected random error\added[id=A]{s}. However, \replaced[id=V]{given}{in regard to} the small number of observations per transit ($\approx7$) and \deleted[id=V]{to} the relative magnitude between random and systematic errors \replaced[id=P]{(for G>13, the random error is at least 40 times greater than the systematic one in the AC direction)}{\added[id=A]{(MAYBE YOU CAN SPECIFY THIS RATIO ?)}}, it is clear \replaced[id=A]{that}{than} in our case, the mean values are strongly \replaced[id=V]{influenced}{governed} by random errors and cannot be taken as accurate estimations of the systematic ones. Typically, for $n$ values \added[id=A]{issued} from a Gaussian distribution with a variance $\sigma^2$, the standard deviation of the mean \replaced[id=V]{is given by}{will be} $\sigma/\sqrt(n)$. In the AC direction, {\obf{where the}} the random error roughly equals to 300 mas and {\obf{where there are an}} average of 7 observations par transit, we find a standard error of about 110 mas on the mean, which is much more  than the expected systematic error (up to 10 mas). This could clearly be seen on Fig \ref{transit1}, looking at the mean residuals per transit in the AC direction, where a value of about 100 mas is obtained for G>13. The transition at G=13 is another consequence of the strong dependenc\replaced[id=A]{y}{e} of the mean \added[id=V]{residuals} with respect to the random error\added[id=A]{s} as no transition is expected in the systematic \added[id=A]{errors}. Finally, the mean \added[id=V]{residuals} per transit cannot be taken as a good estimator of the systematic error in the AC direction as it provides a value with a \added[id=A]{uncertainty} \deleted[id=P]{sigma} bigger than the expected value itself \replaced[id=A]{($S/N < 1$)}{(signal-to-noise ratio smaller than one)}. The situation is \deleted[id=A]{a little bit} different in the AL direction\replaced[id=V]{. In this direction}{ as}\added[id=P]{, for G<17, }the random error is expected to be \replaced[id=P]{smaller}{of the same order \deleted[id=A]{of magnitude}} than the systematic \replaced[id=V]{error}{one} \replaced[id=P]{so that the mean residuals per transit give an estimation of the systematic error with $S/N > \sqrt{N}$.}{, \replaced[id=A]{with $S/N \approx \sqrt{N}$}{such as -roughly- a signal-to-noise  ratio (SNR) of about $\sqrt(N)=\sqrt(7)=2.6$ is expected \added[id=A]{,when considering an average number of observations per transit of about 7}}}\added[id=A]{ For an average number of observation per transit of about 7, the $S/N > 2.6$ } . Thus, the top row of Fig \ref{transit1} \replaced[id=V]{demonstrates}{confirms} the existence of a systematic error smaller than 1 mas in the AL direction, and the bottom figures of Fig \ref{transit1} \replaced[id=V]{shows}{demonstrates} the clear increment of the AL random error\added[id=A]{s} with the magnitude\replaced[id=A]{, t}{. T}he submilliarcsecond accuracy \replaced[id=A]{being}{is} reached for G<17.

\replaced[id=V]{Finally, the correlation between the estimated systematic errors in AL and AC directions must be taken into account in the analysis of the post-fit residuals. The approach discussed in this paper \added[id=P]{and used by \cite{2018A&Atest}} was insufficient, since a rigorous estimation of systematic bias and debiased residuals was required to compare their values to the expected random and systematic errors.}{Finally, as a careful analysis of the post-fit residuals \replaced[id=A]{accounts also for}{must also take into account} the correlation\added[id=A]{s} between the estimated systematic\added[id=A]{s} in AL and AC direction\added[id=A]{s}, the \replaced[id=A]{approach presented previously and used by \ref{xxx}}{presented approach} is insufficient and a rigorous estimation of both systematic bias and debiased residuals is necessary in order to compare their values \replaced[id=A]{with}{to} the expected random and systematic errors.}

\begin{table}
\caption{Means and standard deviations of the Gaia Post-fit residuals in the AC/AL directions.}
\begin{tabular}{>{\bfseries}ccccc} 
    \cmidrule(l){2-5}
    & \multicolumn{2}{c}{AL}& \multicolumn{2}{c}{AC} \\
   \cmidrule(l){2-3}
   \cmidrule(l){4-5}
    & mean & std & mean & std  \\
    & mas & mas & mas & mas \\
    \cmidrule(l){2-2}
    \cmidrule(l){3-3}
    \cmidrule(l){4-4}
    \cmidrule(l){5-5}
    This work & 0.08 & 2.12 & 17.1 & 294.5   \\
    Gaia Collaboration et al (2018b) & 0.05 & 2.14 & - & -  \\
    \cmidrule(l){2-3}
   \cmidrule(l){4-5}
\end{tabular}
\label{tablecompare}
\end{table}

\section{Post-fit analysis and comparisons with radar observations}
\label{postfit}

\subsection{Method}
\label{postfit1}
\subsubsection{Post-fit analysis}

\added[id=V]{To check the accuracy of the orbits obtained after the adjustment with Gaia observations, we carefully analyse the post-fit residuals including the estimation of the Gaia systematic errors.}
\deleted[id=P]{\added[id=P]{In order to validate the orbits obtained after the adjustment \added[id=A]{to} \deleted[id=P]{on} Gaia observations \added[id=A]{following the procedure described in sect. \ref{fit}}, we decide\added[id=P]{d} to perform a careful analysis of their post-fit residuals including an estimation of \added[id=A]{the} \deleted[id=P]{their} Gaia systematic error\added[id=A]{s}.}
\added[id=V]{We initiated the analysis using orbits derived from Gaia observations with a dynamical model consisted of 16 perturbers, and weighting schema based on the random error.}}
We estimate the systematic bias $\beta_i$ by minimizing the Eq 11 such as:

\begin{equation}\label{newS}
S=\sum_{i=1}^{n} \sum_{j=1}^{m_i} (\xi_{ij} - \beta_i)^T W_{ij} (\xi_{ij}-\beta_i)+\sum_{i=1}^{n}\beta_i^T V_i \beta_i
\end{equation}

\noindent $n$ i\replaced[id=V]{s}{n} the number of transits for the considered object, $m_i$ is the number of \added[id=A]{$(\alpha,\delta)$} \deleted[id=P]{(ra,dec)} observation\added[id=A]{s} in the transit $i$, $\xi_{ij}$ is the $j$th residual\deleted[id=A]{s} of the transit $i$, $W_{ij}$ is the random weight matrix for the $j$th observation of the $i$th transit, taken as the inverse of the random covariance matrix\replaced[id=V]{,}{.} $\beta_i$ is the adjusted systematic error for the transit $i$\added[id=V]{,}  and $V_i$ its weight matrix taken as the inverse of the systematic covariance matrix provided by the DPAC.
\added[id=P]{The second term of the sum in Eq \ref{newS} tends to minimize the estimated systematic bias according to the DPAC covariance matrices, whereas the first term of the sum tends to reduce Gaia debiased residuals in accordance with the  announced random errors. Therefore, the minimization of Eq \ref{newS} leads to an optimization of the so-called "Bias variance trade-off".}

\added[id=V]{ After convergence achieved and systematic biases estimated, we examine the three most relevant $\chi^2$ values.} The first one accounts for the systematic errors by considering the sum:

\begin{equation}\label{khi1}
{\chi}^2_{sys}=\sum_{i=1}^{n}\beta_i^T V_i \beta_i
\end{equation}

\noindent whereas the second and the third \added[id=A]{$\chi^2$} \deleted[id=P]{one} deal with the random errors respectively in the AL and AC directions.{\obf{${\chi}^2_{sys}$} measures the compatibility of the estimated systematics bias with the one given by DPAC.}
We \deleted[id=P]{decide to}\deleted[id=V]{preliminary}rotat\added[id=P]{ed} each \added[id=A]{($\alpha,\delta$)} \deleted[id=P]{(rad,dec)} highly correlated Gaia observations into the (AL,AC) plane where \deleted[id=V]{the} longitudinal and cross directions are essentially non correlated and \replaced[id=A]{can then}{then can} be separated into two independent batches. \added[id=P]{The same transformation is applied to} $\beta$ and\deleted[id=V]{to the} Gaia random covariance matrix \replaced[id=V]{to create}{, leading to the creation of}  two diagonal and independent matrices with diagonal terms equal to \deleted[id=P]{espectivly} $\sigma^2[AL]$ and $\sigma^2[AC]$ \added[id=V]{respectively}. We indicate with $\xi[AL]$ and $\beta[AL]$ ((respectively $\xi[AC]$ and $\beta[AC]$)) the part of the residuals and of the bias related to the AL axis (respectively AC). \deleted[id=A]{Then t}Two separated ${\chi}^2$ can \added[id=A]{then} be considered:

\begin{equation}\label{khi2}
{\chi}^2[AL]_{random}=\sum_{i=1}^{n} \sum_{j=1}^{m_i} \frac{(\xi[AL]_{ij} -  \beta[AL]_i)^2}{(\sigma^2[AL]_{ij})}
\end{equation}

\begin{equation}\label{khi3}
{\chi}^2[AC]_{random}=\sum_{i=1}^{n} \sum_{j=1}^{m_i} \frac{(\xi[AC]_{ij} -  \beta[AC]_i)^2}{(\sigma^2[AC]_{ij})}
\end{equation}

\noindent \added[id=A]{The} \deleted[id=P]{Then, the} goodness of the fit is  evaluated using two statistical values derived from the obtained ${\chi}^2$ (Eq \ref{khi1}, \ref{khi2}, and \ref{khi3}): the Survival rate \replaced[id=A]{(SVR)}{(SVR)} and the Standard error of the regression (SER). 
The Survival rate is \deleted[id=P]{defined as} the probability that a real-valued random variable following the ${\chi}^2$ statistic\added[id=A]{s} takes a value \added[id=A]{greater} \deleted[id=P]{bigger} than the obtained ${\chi}^2$. It is computed using the following formula:

\begin{equation}
   SVR({\chi}^2,k) = 1 - F({\chi}^2,k)
\end{equation}

\noindent where $F$ is the cumulative distribution function of a ${\chi}^2$ statistic\added[id=A]{s} with $k$ degrees of freedom:

\begin{equation}
   F({\chi}^2,k) = \frac{\gamma(\frac{k}{2},\frac{{\chi}^2}{2} )}{\Gamma(\frac{k}{2})}
\end{equation}

\noindent $\Gamma$ is the Gamma function and $\gamma$ the lower incomplete one\deleted[id=A]{(REFERENCE ?)}. Usually, for large k and a residual in perfect accordance with the corresponding standard deviation, ${\chi}^2$=k and SVR=50\% \deleted[id=A]{(REFERENCE ?)}. \deleted[id=P]{In least-square analysis, it is generally admitted that a solution with a survival rate greater than 0.3 \added[id=A]{$\%$} \deleted[id=P]{percent} cannot be rejected \deleted[id=A]{(REFERENCE ?)}}

The standard error of the regression (SER) is defined as the multiplicative factor that must be applied to the expected error\deleted[id=V]{in order} to obtain a SVR of 50\% for the considered residual\added[id=A]{s}\deleted[id=A]{(REFERENCE ?)}. For large $k$ it could be simply computed as the squared root of the reduced ${\chi}^2$. A value greater than \added[id=A]{1} \deleted[id=P]{one} indicate\added[id=A]{s} that the error is underestimated (or the residual \added[id=A]{is} too big) whereas a SER smaller than 1 indicates a SVR bigger than 50 $\%$ with a probable overestimation of the error. {\obf{In the case of SVR estimated for the systematic bias with ${\chi}^2_{sys}$}, a SVR bigger than 50$\%$ indicates that the estimated bias is consistent with the one given by DPAC when a SVR close to 0, indicates that the estimated systematic bias has 0 probability of being explained by the DPAC value. In this case, the systematic bias may have absorbed some unmodeled signature.}

\subsubsection{Comparisons with radar  {\obf{ranging}} observations}

\replaced[id=V]{T}{In order t}o validate the quality of the obtained orbits, we also extend\deleted[id=A]{ed} our dataset with \deleted[id=V]{available} radar range observations \replaced[id=V]{available at}{delivered by the JPL} \url{https://ssd.jpl.nasa.gov/?radar}\deleted[id=V]{) where more than 3730 ground-based radar \replaced[id=A]{ranging}{delay} and Doppler measurements are \added[id=A]{released} \deleted[id=P]{delivered} for more than 900 objects, mostly NEA}.\replaced[id=V]{ Spanning from 1993 to 2019, a total of 141 radar {\obf{ranging}} observations were available for 23 asteroids \added[id=P]{also observed by Gaia and }listed in Table \ref{table0}  }{ For our 23 selected objects, \added[id=A]{we have} \deleted[id=P]{it represents} 141 radar \added[id=A]{range} \deleted[id=P]{delay} observations, from 1983 to 2019}. The choice of radar {\obf{ranging}} observations \added[id=A]{in comparison with} \deleted[id=P]{compared to} a more abundant database \added[id=A]{such as optical angular ground-based observations}\deleted[id=P]{in the plane sky)} is justified for 2 reasons: \added[id=A]{f}\deleted[id=P]{F}irst, we need to compare observations of equivalent \added[id=A]{accuracy to GAIA DR2} \deleted[id=P]{and high accuracy} in order to detect potential \replaced[id=A]{inconsistencies} {misalignments} between the two databases. The radar observations bring also the benefit of providing out-of-plane and distance informations \added[id=A]{that are complementary to Gaia angular measurements $(\alpha,\delta)$}\deleted[id=P]{compared to the angular\added[id=A]{$\alpha$}\deleted[id=P]{ra}and Dec}. Second\deleted[id=V]{ly}, at this level of accuracy, the use of optical ground-based observations (especially those obtained before 2000) is \replaced[id=V]{challenging}{\added[id=A]{complex}} \deleted[id=P]{very tricky} as they are\added[id=V]{ usually} very heterogeneous (different observatories, periods, star-catalogs etc) \replaced[id=V]{with poorly known accuracies\added[id=P]{(see for example \cite{CARPINO2003248, 2010Icar..210..158C, 2015Icar..245...94F, 2017Icar..296..139V}).}}{and their accurac\added[id=A]{ies}\deleted[id=P]{y} are not always well known \added[id=A]{(REFERENCE)}} \added[id=A]{A}\deleted[id=P]{Then, a} rigorous treatment of such a database is \replaced[id=V]{beyond}{behind} the scope of this study\deleted[id=P]{,}\added[id=A]{.}

\added[id=P]{We adjust the orbits of the 23 objects of Table \ref{table0} on both Gaia and radar measurements. }
 As we are only working with few individual cases \deleted[id=P]{\added[id=A]{(\added[id=P]{i.e the} 23 objects \deleted[id=P]{were} observed both with GAIA and radar measurements)}} known as perturbed objects, it is relevant to consider them as independent objects from the fit point-of-view. The perturbations induced by the $p$ perturbers are still included in the dynamical modeling but their initial conditions are not re-estimated as they were already fitted with Gaia observations in section \ref{adjustment}.
 
 It is equivalent to not include the perturbing matrices $P_i$ in Eq. \ref{fullmatrix} and to work with $J_i$ matrices extended \added[id=A]{to} \deleted[id=P]{with} radar observations. Therefore, for each of the selected asteroid, the classical least square solution is found using the following combined matrices:

\begin{equation}
W
=
\begin{bmatrix}
W_{Gaia} &0  \\
0& W_{radar} \\
\end{bmatrix}
\end{equation}

\begin{equation}
J
=
\begin{bmatrix}
J_{Gaia}  \\
J_{radar} \\
\end{bmatrix}
\end{equation}

\begin{equation}
\xi
=
\begin{bmatrix}
\xi_{Gaia}  \\
\xi_{radar} \\
\end{bmatrix}
\end{equation}

\noindent where $J_{radar}$ and $J_{Gaia}$ are the Jacobian matrices accounting respectively for the radar and Gaia observations\replaced[id=V]{,}{.} $\xi$ is the \deleted[id=V]{whole} residual \replaced[id=V]{matrix that includes}{including} Gaia and radar \replaced[id=V]{residuals}{ones}. $W_{Gaia}$ is the weight matrix for the Gaia observations. \replaced[id=V]{ As in}{Similar\deleted[id=V]{ly} to} Section \ref{fit}, $W_{Gaia}$ is \replaced[id=V]{computed}{taken} as the inverse of the random covariance matrix provided by the DPAC. $W_{radar}$, the radar weight matrix, \replaced[id=V]{is computed as}{a diagonal matrix where each term is} the inverse of the \replaced[id=V]{formal}{estimated} variance errors of the related observations. This error is provided in the JPL file as a standard-error on the round-trip delay. Those values are conservative and also {\obf{account for}} the uncertaint\added[id=A]{ies}\deleted[id=P]{y} on the asteroid shapes. {\obf{Uncertainties range from few kilometers for the older \added[id=A]{observations} to tens of meters for the most recent.}}  

\replaced[id=P]{Finally a}{A} \deleted[id=V]{specific} ${\chi}^2$ \added[id=A]{is}\deleted[id=P]{was} computed for the radar residuals, as well  as a \replaced[id=V]{SVR}{survival rate} and a \replaced[id=P]{SER}{ standard error of the regression} \added[id=A]{using}:

\begin{equation}
{\chi}^2[radar]=\xi_{radar}^T W_{radar} \xi_{radar}
\end{equation}

\begin{table}
\caption{Gaia and radar datasets for the 23 selected objects. Each Gaia observation accounts for a pair \replaced[id=A]{($\alpha$,$\delta$)}{(ra,dec)}. The orbital coverage is the ratio between the Gaia observation period and the orbital one.}

\begin{tabular}{|>{\bfseries}c|c|c|c|p{3.5cm}|>{\centering\arraybackslash}m{1.cm}|>{\centering\arraybackslash}m{1.cm}|} 
   \hline
    \multirow{2}*{\diagbox[width=2.2cm,height=2.7\line]{Asteroid:}{properties:}} & \multicolumn{2}{m{1.8cm}|}{\makecell{ Number of \\observations:}} & \makecell{Gaia Orbital\\coverage} & \makecell{Radar Observations epoch } \\
   \cline{2-5}
   & Radar & Gaia & [ratio] & \makecell{[year]} \\
   \hline
    53 & 1 & 204 & 0.35 & 2002 \\
   \hline
    105 & 2 & 116 & 0.40 & 1988 \\
   \hline
    216 & 1 & 143 & 0.28 & 1999 \\
   \hline
    253 & 1 & 182 & 0.39 & 2001 \\
   \hline
    393 & 1 & 122 & 0.37 & 2000 \\
   \hline
    433 & 4 & 52 & 0.88 & 2012, 2019 \\
   \hline
    654 & 12 & 55 & 0.41 & 1988, 2002\\
   \hline
    1620 & 3 & 82 & 1.16 & 1983, 1994 \\
   \hline
    1685 & 7 & 102 & 0.78 & 1988, 2012, 2016 \\
   \hline
    2062 & 5 & 69 & 1.64 & 2012, 2013, 2014, 2015\\
   \hline
    2063 & 7 & 22 & 1.21 & 1996, 2012, 2015\\
   \hline
    2100 & 8 & 33 & 1.0 & 2000, 2003, 2016, 2019\\
   \hline
    3103 & 1 & 65 & 0.99 & 1996 \\
   \hline
    3200 & 8 & 118 & 1.03 & 2007, 2017\\
   \hline
    4183 & 1 & 122 & 0.34 & 2000\\
   \hline
    4769 & 8 & 117 & 0.79 & 1989, 2012 \\
   \hline
    7889 & 2 & 45 & 0.4 & 2005\\
   \hline
    10115 & 10 & 103 & 0.71 & 1999\\
   \hline
    11066 & 4 & 117 & 0.49 & 2004\\
   \hline
    16834 & 1 & 41 & 0.79 & 2000\\
   \hline
    66391 & 37 & 24 & 2.04 & 2001, 2019\\
   \hline
    68216 & 5 & 24 & 0.39 & 2009\\
   \hline
    68950 & 12 & 48 & 1.4 & 2003, 2006, 2016, 2019\\
   \hline
\end{tabular}
\label{table0}
\end{table}

\subsection{Results}
\label{postfit2}

\begin{table}
\caption{Gaia post-fit residuals based on \replaced[id=V]{Gaia data alone}{\added[id=A]{Gaia observations only}}\deleted[id=P]{Gaia alone}. The results are based on a dynamical model with 16 perturbers and a weighting schema based on random error only \added[id=A]{(similar to \cite{2018A&Atest})}.}
\begin{tabular}{|>{\bfseries}c|c|c|c|c|c|c||c|c|c|>{\centering\arraybackslash}m{1.cm}|>{\centering\arraybackslash}m{1.cm}}
   \hline
    \multirow{4}*{\diagbox[width=2cm,height=4.2\line]{Asteroid:}{properties:}} & \multicolumn{6}{c||}{\makecell{Gaia residuals:}} & \multicolumn{3}{c|}{\makecell{Radar passthrough:}} \\
    \cline{2-10}
   & \multicolumn{2}{m{1.8cm}|}{\makecell{AL}} & \multicolumn{2}{m{1.8cm}|}{\makecell{AC}} &
   \multicolumn{2}{m{1.8cm}||}{\makecell{Systematic}} &
   \multicolumn{3}{c|}{\makecell{One-way}} \\
   \cline{2-10}
    & SER & SVR & SER & SVR &  SER & SVR & SER & SVR & RMSE \\
    & - & $\%$ & - & $\%$ &  - & $\%$ &  - & $\%$ & km\\
   \hline
    53 & 0.98 & 100 & 0.71 & 100 & 0.22 & 100  & 15.2 &0 & 153.6 \\
    \hline
    105  & 0.94 & 82 & 0.52 & 100 & 0.18 & 100  & 2.6 & 1 & 45\\
    \hline
    216  & 0.87 & 98.9 & 0.39 & 100 & 0.41 & 100  & 14.6 &0& 73.6\\
    \hline
    253  & 0.78 & 100 & 0.27 & 100 & 0.14 & 100  & 92.7 &0 & 562\\
    \hline
    393  & 0.94 & 82.5 & 0.4 & 100 & 0.19 & 100 & 18&0 & 200.6 \\
    \hline
    433  & 1.08 & 21.5 & 0.67 & 100 & 0.1 & 100 & 11.9&0 & 22.54 \\
    \hline
    654 & 1.14 & 7.5 & 0.35 & 100 & 0.26 & 100 & 64.2&0 & 1345\\
    \hline
    1620 & 1.23 & 0.23 & 0.83 & 98.6 & 0.43 & 100 & 438.4&0 & 224.7\\
    \hline
    1685  & 0.97 & 65.1 & 0.57 & 100 & 0.36 & 100 & 24.4&0 & 69.7\\
    \hline
    2062  & 1.18 & 2.16 & 1.01 & 43.6 & 0.26 & 100 & 74&0 & 27.4 \\
    \hline
    2063 & 0.66 & 99.2 & 0.53 & 100 & 0.21 & 100 & 4389&0 & 1527\\
    \hline
    2100  & 1.01 & 47.0 & 0.25 & 100 & 0.03 & 100 & 2519&0 & 1310\\
    \hline
    3103  & 0.87 & 92.9 & 1.01 & 44.3 & 0.19 & 100& 51&0 & 51.6\\
    \hline
    3200  & 1.1 & 7.1 & 0.44 & 100 & 2.39 & 0& 2465&0 & 1412\\
    \hline
    4183  & 1.04 & 27.3 & 0.52 & 100 & 0.28 & 100 & 3466&0 & 1401\\
    \hline
    4769  & 1.03 & 31.8 & 0.94 & 80.3 & 0.31 & 100 & 2695&0 & 339\\
    \hline
    7889  & 1.04 & 36.3 & 0.57 & 100 & 0.02 & 100 & 4770&0 & 2381\\
    \hline
    10115  & 0.95 & 75 & 0.84 & 99.1 & 0.41 & 100 & 164336&0 & 20947\\
    \hline
    11066  & 0.94 & 83.6 & 0.65 & 100 & 0.21 & 100 & 175.4&0 & 63.12\\
    \hline
    16834  & 0.98 & 56.3 & 0.64 & 100 & 0.08 & 100& 310.8&0 & 377.1\\
    \hline
    66391  & 0.84 & 87.8 & 0.6 & 99.9 & 0.28 & 100& 9251&0 &482\\
    \hline
    68216  & 1.44 & 0.23 & 0.64 & 99.6 & 0.02 & 100& 2787&0 & 758\\
    \hline
    68950 & 0.94 & 72 & 1 & 49.7 & 0.26 & 100 & 4261&0 & 1992\\
    \hline
\end{tabular}
\label{table1}
\end{table}

Table \ref{table1} shows the Gaia residual\deleted[id=P]{s} post-fit analysis for the solution adjusted on \added[id=A]{Gaia observations only}\added[id=P]{, obtained in Section \ref{adjustment}} \deleted[id=P]{Gaia alone}. 
\replaced[id=V]{Among 23 asteroids, only one (3200 Phaeton) was rejected due to its zero systematic SVR (in practice, <0.01$\%$) with SER>2. Phaeton extremely elongated and chaotic orbit \citep{2019SoSyR..53..215G} could explain these extreme values.}{We fist notice that only one asteroid (3200 Phaeton) out of 23 must be rejected, as its systematic SVR is equal to zero (in practice <0.01$\%$) with a SER greater than 2. However, as (3200) is known for having an extremely elongated and chaotic orbit \citep{2019SoSyR..53..215G}, a lack in our dynamical modeling could be invoked at this point \added[id=A]{to explain such bad statistics}.}  For (1620), (68216) and (2062), the SVR of 0.23 $\%$, 0.23$\%$ and 2.16$\%$ respectively in the  AL direction are suspicious and indicate less than 5$\%$ of chance that the values of their AL residuals are only due to Gaia AL random errors. For all other objects, the SVR\added[id=A]{s} \replaced[id=A]{are}{is} greater than 7$\%$ \added[id=A]{for} \deleted[id=P]{on} all the \replaced[id=A]{directions}{axes}, confirming the goodness of the fit obtained in Section \ref{adjustment}. We also \replaced[id=V]{check}{confirmed} that most of the constraint is supported by the AL direction, as the SVR \replaced[id=A]{for}{on} AC remains considerably larger (\added[id=A]{20} \deleted[id=P]{eleven} asteroids out of 23 have an AL SVR greater than 80$\%$).
\replaced[id=V]{Based on the analysis of 23 objects, it appears that the use of a dynamical model with 16 perturbers and a weight matrix based on random error is sufficient to adjust the asteroid orbits using Gaia observations only.
However, because Gaia DR2 only covers 22 months of observations, the success of the adjustment is not sufficient for concluding on the reliability of the obtained orbit and its compatibility with other available observations.}{Finally, based on the analysis of those 23 objects, the use of a dynamical model with 16 perturbers and a weight matrix based on \added[id=A]{only} random error \deleted[id=A]{only}, seems to be sufficient \added[id=A]{for adjusting} \deleted[id=P]{in order to adjust} the asteroid trajectories on \added[id=A]{Gaia observations only} \deleted[id=P]{Gaia alone} \added[id=A]{and at the level of the GAIA expected accuracy}. However, as Gaia-DR2 \deleted[id=P]{only covers} \added[id=A]{time coverage is only} 22 months, the success of the adjustment is not a sufficient argument to conclude to the reliability of the obtained orbit\added[id=A]{s} and to \added[id=A]{the} \deleted[id=P]{its} compatibility with other available observations. 
 the data does not provide sufficient evidence to conclude that the observed data differs from the expected.}

\begin{table}
\caption{Analysis of the Gaia post-fit residuals in case of an adjustement on both Gaia and radar, using a dynamical model with 16 perturbers and a weighting schema based on random error only. }
\begin{tabular}{|>{\bfseries}c|c|c|c|c|c|c|c|c||c|>{\centering\arraybackslash}m{1.cm}|>{\centering\arraybackslash}m{1.cm}|} 
   \hline
    \multirow{5}*{\diagbox[width=2cm,height=5.3\line]{Asteroid:}{properties:}} & \multicolumn{9}{c|}{\makecell{residuals:}} \\
    \cline{2-10}
     & \multicolumn{6}{c|}{\makecell{Gaia}} & \multicolumn{3}{c|}{\makecell{Radar}} \\
   \cline{2-10}
   & \multicolumn{2}{m{1.8cm}|}{\makecell{AL}} & \multicolumn{2}{m{1.8cm}|}{\makecell{AC}} &
   \multicolumn{2}{m{1.8cm}|}{\makecell{Systematic}} &\multicolumn{3}{c|}{\makecell{One-way}} \\
   \cline{2-10}
    & SER & SVR & SER & SVR &  SER & SVR & SER & SVR & RMSE \\
    & - & $\%$ & - & $\%$ &  - & $\%$ &  - & $\%$ & km \\
   \hline
    53 & 0.98 & 64.6 & 0.71 & 100 & 0.22 & 100 & 4.92 & 0.09 & 49.7 \\
    \hline
    105 & 0.94 & 82 & 0.51 & 100 & 0.18 & 100 & 2.52 & 1.23 & 44 \\
    \hline
    216 & 0.87 & 98.9 & 0.39 & 100 & 0.41 & 100 & 1.7 & 25.1 & 8.6 \\
    \hline
    253 & 0.78 & 100 & 0.27 & 100 & 0.16 & 100 & 7.45 & 0 & 45.2 \\
    \hline
    393 & 0.94 & 82.5 & 0.4 & 100 & 0.2 & 100 & 8.67 & 0 & 96.5 \\
    \hline
    433 & 1.08 & 21.5 & 0.67 & 100 & 0.11 & 100 & 5.9 & 0 & 15.2 \\
    \hline
    654 & 1.14 & 7.45 & 0.35 & 100 & 0.29 & 100 & 1.21 & 16.2 & 43.7\\
    \hline
    1620 & 1.24 & 0.17 & 0.83 & 98.8 & 1.6 & 0 & 165.2 & 0 & 494.8 \\
    \hline
    1685 & 0.97 & 65.2 & 0.57 & 100 & 0.34 & 100 & 7.96 & 0 & 15.2  \\
    \hline
    2062  & 1.18 & 1.9 & 1 & 52.3 & 0.52 & 100 & 4.52 & 0 & 1.39  \\
    \hline
    2063 & 0.78 & 93.1 & 0.88 & 78.3 & 1.18 & 18.7 & 15.1 & 0 & 5.7  \\
    \hline
    2100 & 1.38 & 0.17 & 0.59 & 100 & 3.83 & 0 & 60.7 & 0 & 50.3  \\
    \hline
    3103 & 0.88 & 92.3 & 1.01 & 45.7 & 0.24 & 100 & 0.33 & 82.1 & 0.34  \\
    \hline
    3200 & 1.14 & 2.11 & 0.44 & 100 & 4.04 & 0 & 132.8 & 0 & 96.3\\
    \hline
    4183 & 1.04 & 28.1 & 0.52 & 100 & 0.24 & 100 & 0.02 & 99.1 & 0.01 \\
    \hline
    4769 & 4.28 & 0 & 0.93 & 86.5 & 10 & 0 & 162.2 & 0 & 148.9 \\
    \hline
    7889 & 1.04 & 36.2 & 0.58 & 100 & 0.02 & 100 & 0.39 & 89.8 & 0.2 \\
    \hline
    10115 & 0.93 & 83.9 & 1.03 & 35.3 & 0.35 & 100 & 0.24 & 100 & 0.32 \\
    \hline
    11066 &  0.95 & 79.4 & 0.64 & 100 & 0.67 & 99.9 & 3.32 & 0 & 1.14 \\
    \hline
    16834 & 0.98 & 56 & 0.63 & 100 & 0.18 & 100 & 0.03 & 98.6 & 0.03 \\
    \hline
    66391 & 13.12 & 0 & 2.44 & 0 & 51.5 & 0 & 79.3 & 0 & 8.9 \\
    \hline
    68216 & 1.44 & 0.23 & 0.63 & 99.7 & 0.23 & 100 & 0.48 & 96.1 & 0.21 \\
    \hline
    68950 & 1.26 & 0.79 & 1.01 & 44.3 & 3.8 & 0 & 29.9 & 0 & 16.2 \\
    \hline
\end{tabular}
\label{table2}
\end{table}

The results of the orbital adjustment on both Gaia and radar \replaced[id=A]{are}{is} presented on Table \ref{table2}. The study of the SVR shows that for \added[id=A]{13} out of 23 objects ((253), (393), (433), (1620), (1685), (2062), (2063), (2100), (3200), (4769), (11066), (66391) and (68950)),  the simple combination of radar and Gaia \replaced[id=A]{does not}{don't} lead to \replaced[id=A]{acceptable}{convenient} residuals. For all these 13 \added[id=A]{objects}, the radar residuals \added[id=A]{are}\deleted[id=P]{is} greater than the expected error (SER>3 and up to 165) and for 6 of the mentioned 13 ((1620), (2100), (3200), (4769), (66391) and (68950)), the obtained systematic error is also bigger than expected, leading to a SVR of 0 $\%$. \added[id=A]{The asteroids} (53) and (105) must probably be also rejected as their radar SVR fall to 0.09 $\%$ and 1.23$\%$ respectively.
For 7 asteroids out of 23 ((216), (654), (3103), (4183), (7889), (10115), (16834)), the adjustment on both Gaia and radar measurements is a success as we obtain radar SVR>15$\%$ and Gaia SVR>7$\%$. Moreover, for all of these 7 asteroids, the Gaia residuals were not impacted by the extension of the dataset to radar observations, as the  Gaia SVRs are roughly the same on Table \ref{table1} and Table \ref{table2}. The case of (68216) is ambiguous as the Gaia AL SVR is small (SVR=0.23$\%$) but constant whether or not the radar measurements are included into the orbital adjustment. Our capability to obtain good radar residuals for (68216) without degrading the Gaia ones could indicates that, in this case, the small Gaia SVR is due more to Gaia data themselves rather than our dynamical model.    

Finally, even if the orbital adjustment on \added[id=A]{Gaia observations only} \deleted[id=P]{Gaia alone} could be considered as a success (Table \ref{table1}), it is clearly not the case when considering both Gaia and radar measurements (Table \ref{table2}). In order to understand if the apparent incompatibility between Gaia and radar is due to the data \added[id=A]{themselves} \deleted[id=P]{themself} or to our \deleted[id=A]{simulation and} adjustment method, we explore\deleted[id=P]{d} in the next section the impact of two hypotheses: the use of only 16 perturbers in the dynamical model, and the neglect of the systematic error\added[id=A]{s} in the weighting schema.

\section{Improving the \replaced[id=A]{procedure of the asteroid orbit determination}{asteroid orbit}}
\label{solution}

\subsection{\replaced[id=A]{Modification in the dynamical modeling:}{Impact of The dynamical model:} \replaced[id=A]{Increase of}{increasing} the number of massive perturbers}

\label{solution1}

In INPOP19a, the number of massive asteroids taken into account during the orbital \added[id=A]{integration}\deleted[id=P]{determination} is 353 for the planets and 16 for the asteroids. Beyond the comparison with  \cite{2018A&Atest} \added[id=A]{where only 16 perturbers were included in the orbital computation}, this difference \added[id=A]{in the dynamical modeling} can also be explained by practical reasons. For a system of $n$ objects with $p$ perturbers, the total number of interactions $T(n,p)$ to \deleted[id=A]{take into} account \added[ad=A]{for} is:

\begin{equation}
T(n,p)=(n-p) \times p + p \times (p-1)/2
\label{eq:20}
\end{equation}

\noindent \deleted[id=A]{which increase\added[id=A]{s} linearly with $n$ like a concave down quadratic function with $p$.} \replaced[id=A]{Approximately, from Eq. \ref{eq:20}, one can say that}{Roughly}, considering the 14099 perturbed objects of Gaia \added[id=A]{DR2}, \replaced[id=A]{the increase of}{increasing} the number of massive perturbers from 16 to 353 \deleted[id=P]{will} multipl\added[id=A]{ies}\deleted[id=P]{y} the number of interaction\added[id=A]{s} by a factor of 20. As each iteration takes 15 hours, the computational time for one iteration \added[id=A]{passes} \deleted[id=P]{will} roughly \deleted[id=P]{pass} from half a day to more than a week. 
\added[id=A]{Nevertheless}\deleted[id=P]{In the other side}, it is clear that the \deleted[id=A]{long-term} accuracy and the predictive \replaced[id=A]{capability}{power} of our trajectories will increase with the number of perturbers accounted \added[id=A]{for} in the dynamical model. The goodness of the orbital adjustment on both Gaia and radar \added[id=A]{observations} is expected to follow the same \added[id=A]{trend}\deleted[id=P]{tendency}. \added[id=A]{We thus}\deleted[id=P]{Thus, we} perform\deleted[id=A]{ed} a new orbital adjustment for the 23 selected objects (for which the computational time remains reasonable) using the same dynamical model as the one used for planets (353 massive perturbers). {\obf{The asteroid masses are kept fixed and are not fitted in the present work.}} For a sake of comparison, we maintain the same weighting schema and adjustment procedure as the ones used in the previous section. 

\begin{table}
\caption{\deleted[id=A]{Analysis of the} Gaia post-fit residuals in case of an adjustment on both Gaia and radar, using a dynamical model \replaced[id=A]{including}{with} 353 perturbers and a weighting schema based on random error only. }
\begin{tabular}{|>{\bfseries}c|c|c|c|c|c|c|c|c||c|>{\centering\arraybackslash}m{1.cm}|>{\centering\arraybackslash}m{1.cm}|} 
   \hline
    \multirow{5}*{\diagbox[width=2cm,height=5.3\line]{Asteroid:}{properties:}} & \multicolumn{9}{c|}{\makecell{residuals:}} \\
    \cline{2-10}
     & \multicolumn{6}{c|}{\makecell{Gaia}} & \multicolumn{3}{c|}{\makecell{Radar}} \\
   \cline{2-10}
   & \multicolumn{2}{m{1.8cm}|}{\makecell{AL}} & \multicolumn{2}{m{1.8cm}|}{\makecell{AC}} &
   \multicolumn{2}{m{1.8cm}|}{\makecell{Systematic}} &\multicolumn{3}{c|}{\makecell{One-way}} \\
   \cline{2-10}
    & SER & SVR & SER & SVR &  SER & SVR & SER & SVR & RMSE \\
    & - & $\%$ & - & $\%$ &  - & $\%$ &  - & $\%$ & km \\
    \hline
        53 & 0.98 & 64.6 & 0.71 & 100 & 0.22 & 100 & 3.23 & 2.94 & 32.66 \\
    \hline
    105 & 0.94 & 82 & 0.51 & 100 & 0.18 & 100 & 2.35 & 2.15 & 41.12 \\
    \hline
    216 & 0.87 & 98.9 & 0.39 & 100 & 0.42 & 100 & 2.7 & 6.79 & 13.7 \\
    \hline
    253 & 0.78 & 100 & 0.27 & 100 & 0.17 & 100 & 8.1 & 0 & 48.9 \\
    \hline
    393 & 0.94 & 82.5 & 0.4 & 100 & 0.2 & 100 & 5.98 & 0 & 66.6 \\
    \hline
    433 & 1.08 & 21.5 & 0.67 & 100 & 0.11 & 100 & 4.7 & 0 & 12.1 \\
    \hline
    654 & 1.14 & 7.45 & 0.35 & 100 & 0.27 & 100 & 0.99 & 51.1 & 37.9\\
    \hline
    1620 & 1.23 & 0.23 & 0.83 & 98.6 & 0.39 & 100 & 1.17 & 35.8 & 1.93 \\
    \hline
    1685 & 0.97 & 65.2 & 0.57 & 100 & 0.31 & 100 & 3.37 & 0 & 4.59  \\
    \hline
    2062  & 1.18 & 2.02 & 1 & 47.6 & 0.45 & 100 & 1.35 & 16.2 & 0.39  \\
    \hline
    2063 & 0.67 & 99 & 0.92 & 70.5 & 0.27 & 100 & 0.85 & 71.2 & 0.38  \\
    \hline
    2100 & 1.01 & 46.4 & 0.54 & 100 & 0.24 & 100 & 0.75 & 84.1 & 0.58  \\
    \hline
    3103 & 0.88 & 92.6 & 1.01 & 45.7 & 0.26 & 100 & 0.48 & 74.4 & 0.49  \\
    \hline
    3200 & 1.08 & 12.5 & 0.45 & 100 & 0.3 & 100 & 5.02 & 0 & 4.23\\
    \hline
    4183 & 1.04 & 28.4 & 0.52 & 100 & 0.23 & 100 & 0.03 & 98.6 & 0.01 \\
    \hline
    4769 & 1.03 & 29.9 & 0.94 & 80.4 & 0.34 & 100 & 2.1 & 0 & 1.66 \\
    \hline
    7889 & 1.04 & 36.2 & 0.58 & 100 & 0.02 & 100 & 0.21 & 97.1 & 0.1 \\
    \hline
    10115 & 0.93 & 83.5 & 1.04 & 30.1 & 0.35 & 100 & 0.23 & 100 & 0.28 \\
    \hline
    11066 &  0.94 & 83.2 & 0.65 & 100 & 0.26 & 100 & 0.47 & 94.6 & 0.16 \\
    \hline
    16834 & 0.98 & 56 & 0.63 & 100 & 0.17 & 100 & 0.03 & 98.6 & 0.03 \\
    \hline
    66391 & 0.86 & 83.5 & 0.5 & 100 & 0.97 & 55.7 & 2.16 & 0 & 0.2 \\
    \hline
    68216 & 1.44 & 0.23 & 0.65 & 99.5 & 0.13 & 100 & 0.54 & 93.8 & 0.27 \\
    \hline
    68950 & 0.94 & 71.5 & 1.02 & 42.9 & 0.41 & 100 & 2.15 & 0 & 1.21 \\
    \hline
\end{tabular}
\label{table3}
\end{table}

The \added[id=A]{statistics of} post-fit residuals \deleted[id=A]{analysis} after the adjustement on both radar and Gaia \added[id=A]{observations} \replaced[id=A]{are}{is} presented in Table \ref{table3}. Compared to the orbits obtained with \deleted[id=A]{the} 16 perturbers\deleted[id=A]{model}, a clear improvement of the residuals is visible \replaced[id=A]{with}{as} only 8 objects\deleted[id=A]{remain\deleted[id=P]{s}}  with a SVR equals to zero (all on the radar residuals). 13 objects use to have SVRs equal to 0 in the previous case. \deleted[id=A]{5 objects have then seen their residuals improved.}
(2063), (2100) and (11066) exhibit acceptable SVR values (>70$\%$ on both Gaia and radar) while they were rejected with the previous dynamical model (with radar SVR equals to 0$\%$). (1620) and  (2062) still have a small Gaia AL SVR of 0.23$\%$ and 2.02$\%$ respectively, but similar to the ones obtained with the solution fitted only on Gaia (Table \ref{table1}) and with a radar SVR >15$\%$.
The strong impact of the new dynamical model is also visible when comparing the radar statistics with the ones of Table \ref{table2}. For \added[id=A]{19}\deleted[id=P]{nineteen} out of 23 \added[id=A]{objects}, \replaced[id=A]{the increase of the number of perturbers from 16 to 353 leads to a \added[id=A]{significant} reduction of the radar residuals, even when the SVR is \added[id=A]{still} zero.}{
even when the SVR is zero, \added[id=A]{demonstrating that the increase of} \deleted[id=P]{increasing} the number of perturbers from 16 to 353 leads to a \added[id=A]{significant} reduction of the radar residuals.} 
For (3200), the improvement also includes the Gaia residuals with a SVR of 12.5$\%$ in AL, 100$\%$ in AC and 100$\%$ for the systematic bias (respectively 7.1$\%$, 100$\%$ and 0$\%$ in Table \ref{table1}) and confirms that the badness of the adjustment of (3200) on Gaia only (Table \ref{table1}) was caused by a lack in our previous dynamical model. 

Another interesting point is to notice that for all the \deleted[id=A]{8} rejected combinations, the rejection criteria is exclusively encountered \replaced[id=A]{for}{on} the radar residuals as, in the same time, the Gaia ones remain\deleted[id=P]{s} acceptable. This behavior could be typically the sign of an over\added[id=A]{-}weighting of the Gaia observations. In the next section, we study the impact of such an hypothesis by changing the weighting schema of our adjustment in order to take into account the impact of Gaia systematic error\added[id=A]{s} during the \added[id=A]{fit} \deleted[id=P]{orbital determination procedure}.  

\subsection{\replaced[id=A]{Modification in the  weighting schema:}{Impact of the weighting schema:} \replaced[id=A]{Introduction of}{taking} the systematic errors \deleted[id=A]{into account}}
\label{solution2}

In Sect. \ref{postfit1}, we proposed a method to compute the values of the Gaia systematic bias based on the minimization of the squared function (see Eq. \ref{newS}). However, as the systematic bias was evaluated during the post-fit analysis, the asteroid initial conditions were \deleted[id=A]{fixed and still} adjusted on the basis of the squared function of Eq. \ref{S}, using a weight matrix that only account\added[id=A]{s} for random errors. \deleted[id=P]{Therefore, it is}\added[id=A]{It is then} clear that the asteroid initial conditions were not optimized for the point of view of Eq. \ref{newS}. In particular, Eq. \ref{S} \replaced[id=A]{sees}{takes} any increment of the systematic bias at the transit level \replaced[id=A]{as}{for} an increment of the \replaced[id=A]{estimated}{evaluated} random error\added[id=A]{s} of all the  observations in the transit. As a consequence, \replaced[id=A]{the weighting schema is skewed leading}{this skewed weighting schema, leads} \deleted[id=P]{, among other,} to an over-constraint on the Gaia systematic bias, visible in Table \ref{table3} where the trade-off between Gaia and radar residuals always favours the  Gaia \replaced[id=A]{observations}{side}.

There are two methods, mathematically equivalent, to fully \deleted[id=A]{take into} account \added[id=A]{for} both the systematic and the random error\added[id=A]{s in the fit} \deleted[id=P]{ during the orbital adjustment procedure}.
The first one consists \added[id=A]{in adding into the initial covariance matrix}\deleted[id=P]{to add} the Gaia systematic error\added[id=A]{s} in quadratic with the random one \deleted[id=A]{into the covariance matrix}, and then to use this new matrix as the inverse of the weight\deleted[id=A]{ing} matrix $W_{Gaia}$ ( see \cite{BARLOW2021164864}), so that Eq. \ref{S} could be rewritten as:

\begin{equation}
S=\xi_{Gaia}^T. [W_{random+systematic}]. \xi_{Gaia}
\label{modifiedS}
\end{equation} 

\noindent \replaced[id=A]{Eq. \ref{modifiedS} is possible because}{This method is based on the fact than} the random and the systematic error\added[id=A]{s} are independent so that \replaced[id=A]{one can add their variances}{their variance\added[id=A]{s} are additive}. In practice, the 2x2 systematic covariance matrix provided by the DPAC for a transit with \replaced[id=A]{$m$}{$p$} observations, \replaced[id=A]{has to be}{must be} extended to a \replaced[id=A]{$m$x$m$}{$p$x$p$} covariance matrix before its addition to the random error covariance matrix. If we call \added[id=A]{$\sigma_{\alpha_s}$, $\sigma_{\delta_s}$} and $cov_{s}$, the systematic standard errors and the systematic covariance between \added[id=A]{$\alpha$}\deleted[id=P]{ra} and \added[id=A]{$\delta$}\deleted[id=P]{dec}  for a given transit, the \replaced[id=A]{$m$x$m$}{$p$x$p$} systematic covariance matrix accounting for all the observations of the transit will be:

\begin{equation}
COV_{sys}
=
\begin{bmatrix}
\sigma^2_{\alpha_s} &cov_{s}  &\sigma^2_{\alpha_s} &cov_{s} &\sigma^2_{\alpha_s} &...\\
cov_{s}& \sigma^2_{\delta_s} & cov_{s} &\sigma^2_{\delta_s} &cov_{s} &..\\
\sigma^2_{\alpha_s}&cov_{s} & \sigma^2_{\alpha_s} &cov_{s} &\sigma^2_{\alpha_s} &..\\
cov_{s}&\sigma^2_{\delta_s}&cov_{s} &\sigma^2_{\delta_s} &cov_{s} &...\\
\sigma^2_{\alpha_s}&cov_{s}&\sigma^2_{\alpha_s}&cov_{s} & \sigma^2_{\alpha_s} &... \\
\vdots&\vdots&\vdots&\vdots&\vdots & \ddots\\
\end{bmatrix}
\end{equation}

\noindent The presence of non-diagonal terms guarantees that a correlation equal to one is expected - from the point of view of the systematic - between each observation of the same transit, as they are impacted by the same systematic error. 

One of the minor disadvantage of this method, based on the minimization of Eq \ref{modifiedS}, is that it doesn't directly provide the value of the obtained systematic bias,  which still \added[id=A]{has to} \deleted[id=P]{must} be computed following the procedure described in Sect. \ref{postfit}. Another method, mathematically \added[id=A]{similar} \deleted[id=P]{equivalent} \citep{BARLOW2021164864}, consists \added[id=A]{in}\deleted[id=P]{to} introduc\added[id=A]{ing the} systematic bias as \replaced[id=A]{additional}{extra-} parameters in the fitting procedure and to directly minimize Eq. \ref{newS}, in order to properly optimize the so-called bias-variance tradeoff. In the case where radar measurement\added[id=A]{s} are also included in the adjustment, Eq \ref{newS} must be rewritten as:  

\begin{equation}\label{ultimateS}
S=\sum_{i=1}^{n} \sum_{j=1}^{m_i} (\xi_{ij} - \beta_i)^T W_{ij} (\xi_{ij}-\beta_i)+\sum_{i=1}^{n}\beta_i^T V_i \beta_i + \xi_{radar}^T W_{radar} \xi_{radar}
\end{equation}

\noindent The goal of the new least-square problem is now to minimize Eq \ref{ultimateS}. The adjusted parameters are the asteroid initial condition\added[id=A]{s} and the vector $\beta$ of the Gaia systematic bias, which are also include\added[id=A]{d} into the complete residuals:

\begin{equation}
\xi
=
\begin{bmatrix}
\epsilon_{Gaia}  \\
\beta_{sys} \\
\xi_{radar} \\
\end{bmatrix}
\label{newres}
\end{equation}

\noindent $\epsilon_{Gaia}$ is the vector of the debiased residuals $(\xi_{ij} - \beta_i)$ presented in Eq. \ref{ultimateS}, $\beta_{sys}$ is the vector of the estimated bias and $\xi_{radar}$ the radar residuals. The weight\deleted[id=A]{ing} matrix become\added[id=A]{s}:

\begin{equation}
W
=
\begin{bmatrix}
W_{random} &0 & 0  \\
0& W_{sys} & 0 \\
0&0& W_{radar} \\
\end{bmatrix}
\label{newW}
\end{equation}

\noindent and the Jacobian J: 

\begin{equation}
J
=
\begin{bmatrix}
J_{Gaia} & J_{sys}  \\
0 & \mathbb{1}\\
J_{radar} & 0 \\
\end{bmatrix}
\label{newJ}
\end{equation}

\noindent where \added[id=A]{ $\mathbb{1}$ being the unit matrix and} $J_{sys}$ is the matrix of the derivative of $\epsilon_{gaia}$ with respect to $\beta$  such as, for each element ij:


\begin{equation}
J_{{sys}}(ij) = \frac{\partial\epsilon_{i}}{\partial \beta_j} =
    \begin{cases}
      1 & \text{if the observation i belongs to the transit j}\\
      0 & \text{otherwise}
    \end{cases}  
\end{equation}

\begin{table}
\caption{\deleted[id=A]{Analysis of the} Gaia post-fit residuals in case of an adjustement on both Gaia and radar, using a dynamical model \replaced[id=A]{including}{with} 353 perturbers and a weighting schema based on both systematic and random error. }
\begin{tabular}{|>{\bfseries}c|c|c|c|c|c|c|c|c||c|>{\centering\arraybackslash}m{1.cm}|>{\centering\arraybackslash}m{1.cm}|} 
   \hline
    \multirow{5}*{\diagbox[width=2cm,height=5.3\line]{Asteroid:}{properties:}} & \multicolumn{9}{c|}{\makecell{residuals:}} \\
    \cline{2-10}
     & \multicolumn{6}{c|}{\makecell{Gaia}} & \multicolumn{3}{c|}{\makecell{Radar}} \\
   \cline{2-10}
   & \multicolumn{2}{m{1.8cm}|}{\makecell{AL}} & \multicolumn{2}{m{1.8cm}|}{\makecell{AC}} &
   \multicolumn{2}{m{1.8cm}|}{\makecell{Systematic}} &\multicolumn{3}{c|}{\makecell{One-way}} \\
   \cline{2-10}
    & SER & SVR & SER & SVR &  SER & SVR & SER & SVR & RMSE \\
    & - & $\%$ & - & $\%$ &  - & $\%$ &  - & $\%$ & km \\
    \hline
        53 & 0.98 & 64.6 & 0.71 & 100 & 0.22 & 100 & 0.01 & 99.7 & 0.06 \\
    \hline
    105 & 0.94 & 82 & 0.52 & 100 & 0.18 & 100 & 0.78 & 65.9 & 13.6 \\
    \hline
    216 & 0.87 & 98.9 & 0.39 & 100 & 0.41 & 100 & 0 & 99.8 & 0.02 \\
    \hline
    253 & 0.78 & 100 & 0.27 & 100 & 0.17 & 100 & 0.01 & 99.3 & 0.08 \\
    \hline
    393 & 0.94 & 82.5 & 0.4 & 100 & 0.19 & 100 & 0.02 & 99.1 & 0.18 \\
    \hline
    433 & 1.08 & 21.5 & 0.67 & 100 & 0.17 & 100 & 0.02 & 100 & 0.06 \\
    \hline
    654 & 1.14 & 7.45 & 0.35 & 100 & 0.26 & 100 & 0.36 & 100 & 20.9\\
    \hline
    1620 & 1.23 & 0.24 & 0.83 & 98.8 & 0.37 & 100 & 0.63 & 81.3 & 1.74 \\
    \hline
    1685 & 0.97 & 65.5 & 0.57 & 100 & 0.4 & 100 & 0.27 & 100 & 0.4  \\
    \hline
    2062  & 1.18 & 2.04 & 1 & 48.2 & 0.46 & 100 & 0.16 & 100 & 0.08  \\
    \hline
    2063 & 0.67 & 99 & 0.93 & 70 & 0.38 & 100 & 0.44 & 99 & 0.32  \\
    \hline
    2100 & 1.01 & 45.9 & 0.55 & 100 & 0.33 & 100 & 0.61 & 95.1 & 0.45  \\
    \hline
    3103 & 0.87 & 93 & 1.01 & 45 & 0.23 & 100 & 0.02 & 98.9 & 0.02  \\
    \hline
    3200 & 1.08 & 12.5 & 0.45 & 100 & 0.56 & 100 & 0.43 & 99.4 & 0.79 \\
    \hline
    4183 & 1.04 & 28.7 & 0.52 & 100 & 0.2 & 100 & 0 & 100 & 0 \\
    \hline
    4769 & 1.03 & 30.9 & 0.95 & 79.9 & 0.53 & 100 & 0.73 & 86.9 & 0.48 \\
    \hline
    7889 & 1.04 & 36.2 & 0.56 & 100 & 0.15 & 100 & 0.19 & 97.6 & 0.09 \\
    \hline
    10115 & 0.93 & 83.8 & 1.04 & 29.9 & 0.35 & 100 & 0.23 & 100 & 0.28 \\
    \hline
    11066 &  0.94 & 83.8 & 0.65 & 100 & 0.23 & 100 & 0.25 & 99.5 & 0.08 \\
    \hline
    16834 & 0.98 & 56.3 & 0.62 & 100 & 0.19 & 100 & 0 & 100 & 0 \\
    \hline
    66391 & 0.89 & 78.3 & 0.49 & 100 & 1.48 & 1.02 & 0.45 & 100 & 0.09 \\
    \hline
    68216 & 1.44 & 0.23 & 0.65 & 99.5 & 0.14 & 100 & 0.51 & 95.1 & 0.26 \\
    \hline
    68950 & 0.95 & 69 & 1.02 & 43 & 0.68 & 99.2 & 1.01 & 47.3 & 0.51 \\
    \hline
\end{tabular}
\label{table4}
\end{table}

Substituting Eq. \ref{newW}, \ref{newJ}, \ref{newres} into Eq. \ref{gn},  we perform a new orbital adjustment of the \added[id=A]{orbits of the} 23 selected objects \replaced[id=A]{using}{on} both Gaia and radar measurements, \replaced[id=A]{and}{using} the same list of 353 perturbers \replaced[id=A]{as}{than} in Sect. \ref{solution1}.
The result of the post-fit analysis is presented on Table \ref{table4}. We obtain \replaced[id=A]{acceptable}{satisfactory} SVR\added[id=A]{s} for all the 23 studied objects confirming that the change in the weighting schema combined with a more complete dynamical model allow\added[id=A]{s} the orbital adjustment on both Gaia and radar with INPOP19a. 

Two concerns can \replaced[id=A]{raise}{be raised} about the values obtained on Table \ref{table4}. First, the small AL SVR (0.24 $\%$, 2.04 $\%$ and 0.23 $\%$ respectively) obtained for (1620), (2062) and (68216) appear to be suspicious. However, \deleted[id=A]{it is noticeable than} these values remain constant irrespective of the number of perturbers \deleted[]{is 16 or 353 and whether} \added[id=A]{or the introduction of} the systematic error\added[id=A]{s} \deleted[id=A]{were taken \deleted[id=P]{or not} into account \added[id=A]{or not} during the adjustment}. This could \replaced[id=A]{mean}{be the sign} that the misfit is more \added[id=A]{due to} \deleted[id=P]{ coming from} the data \added[id=A]{themselves} \deleted[id=P]{itself} than from the dynamical model \added[id=A]{or the fit}. 
The second remark \added[id=A]{stands for} \deleted[id=P]{concerns} (66391) - a binary NEA - for which the systematic SVR is only 1.02 $\%$, contrasting with values greater than 99 \added[id=A]{$\%$}\deleted[id=P]{percent} obtained for all the others objects. Comparing Table \ref{table2} and Table \ref{table3}, we note that the residuals of (66391) are very sensitive to the change in the dynamical model, including GR (\cite{2017ApJ...845..166V})\deleted[id=P]{, such as it cannot be}\added[id=A]{. We cannot then} exclude\deleted[id=P]{d}  that its relative misfit could be still caused by a lack from this side. In particular, (66391) is known \deleted[id=A]{(see \cite{Greenberg_2020} for example)} for being a candidate to  Yarkosvsky effect detection 
\deleted[id=P]{ -  a small  recoil  force  that  is  due  to  the  emission  of photons in  the  thermal  infrared  from  the  surface  of  SSOs- which causes a drift in the semi-major axis of asteroids} \citep{2000Icar..148..118V,2020AJ....159...92G}. Adding such a non-gravitational perturbation in our dynamical model is behind the scope of this paper but could be an interesting possibility in the perspective of Gaia DR3. 
{\obf{In the case of (1620) and (2062), we note that our modeling did not include Doppler measurements that have been obtained together with or even before the range data. These Doppler observations are crucial for poorly constrained objects as they give an accurate estimation of the asteroid radial velocity which is an important parameter especially for very fast objects. 
For (1620), they were even more Doppler measurements than range observations. For (2062) there are more range observations (6 against 2 Doppler) but the first Doppler observation was obtained 17 years before the range session, increasing the time interval of the fit and consequently the time resolution of the adjustment. A next future improvement will be to include the Doppler observations to our analysis.
}}
\section{Conclusion and perspective}

We perform the orbital determination of 14099 \added[id=A]{asteroids}\deleted[id=P]{SSOs} for which observations have been delivered on the Gaia \added[id=A]{DR2}\deleted[id=P]{data release 2}, \added[id=A]{gathering}\deleted[id=P]{representing} nearly 2 millions of observations, using INPOP19a \added[id=A]{planetary ephemerides} and \replaced[id=A]{a dynamical modeling similar to \cite{2018A&Atest}}{a 16 body perturbing model}. Two new aspects \added[id=P]{were}\deleted[id=P]{have been} investigated \deleted[id=A]{compared to the reference adjustment \citep{2018A&A}}:
\begin{itemize}
\item \added[id=P]{we point out the fact that \added[id=A]{because of} the \replaced[id=A]{SSB}{Solar System barycenter} \added[id=A]{shift}\deleted[id=P]{shifted} between INPOP10e and INPOP19a due to the addition of ten massives TNOs, \deleted[id=P]{such as} the barycentric positions of the Gaia \added[id=A]{satellite}\deleted[id=P]{spacecraft} provided by DR2 \added[id=A]{at J2000} cannot be directly used. Assuming that the geocentric state is obtained with \added[id=A]{ground-based} tracking data, we \replaced[id=A]{then consider}{shift} the Earth-spacecraft vector \added[id=A]{that we shift }from INPOP10e to INPOP19a in order to \added[id=A]{deduce}\deleted[id=P]{recover} the \deleted[id=P]{correct} Gaia position\added[id=A]{s into the INPOP19a frame}. However, to avoid a systematic comparison with INPOP10e when using another ephemerides, we recommend\deleted[id=V]{, if possible,} to provide \replaced[id=A]{for the future Gaia releases,}{in the future} the Gaia \added[id=A]{satellite}\deleted[id=P]{spacecraft} positions in a \replaced[id=V]{minimal}{less} model-depend\deleted[id=P]{e}\added[id=P]{e}nt parametrization \deleted[id=P]{(system of equation, telemetry, geocentric)}}.
\item we \deleted[id=P]{have} develop a convenient \added[id=P]{mathematical} formalism \deleted[id=P]{for the inversion of the least-square} \added[id=P]{for the resolution of the least-square problem}, according for perturbers influences on the perturbed orbits. This formalism will be very useful in the perspective of Gaia DR3 as it presents two advantages:  it is capable of handling small matrices independently from the number of observations. It includes also global parameters such as object masses. \replaced[id=A]{The estimation of global parameters is not possible if one {\obf{adjusts}} the orbits as if they were independent. By introducing such formalism we prepare the venue of the Gaia DR3 and the determination of masses together with the orbit adjustment.}{ which may not be possible if we only invert orbit\added[id=A]{s} as if they were independent.}
\end{itemize}
\added[id=P]{As a result \added[id=A]{and considering only Gaia observations,}} we obtain Gaia post-fit residuals in good accordance with \replaced[id=A]{\cite{2018A&Atest}}{the published ones}, confirming the magnitude dependenc\added[id=A]{y}\deleted[id=P]{e} and the milliarcsecond accuracy level in the AL direction. We also perform a careful post-fit analysis, including the effect of \added[id=A]{the} Gaia systematic error\added[id=A]{s}, for a selection of 23 objects, inferring a good accordance between the obtained residuals and the expected error model announced by Gaia-DR2\\
However, as the Gaia-DR2 \added[id=A]{time coverage is} only \deleted[id=P]{cover} 22 months, we \replaced[id=V]{also perform}{decided to perform} a global orbit adjustment using both Gaia and radar measurements in order to check the validity of the obtained trajector\added[id=A]{ies}\deleted[id=P]{y} and the compatibility of Gaia-DR2 with other dataset\added[id=A]{s}. Using 16 perturbers and a weighting schema \replaced[id=A]{only based on}{based on only} random error\added[id=A]{s}, we \replaced[id=V]{were unable to find  a good fit}{failed} \added[id=A]{in} \deleted[id=P]{to} \added[id=A]{combining} \deleted[id=P]{combine} both Gaia and radar observations as \added[id=A]{13} \deleted[id=P]{thirteen} objects over the mentioned 23 remain with $\chi^2$ greater than expected. \replaced[id=A]{We \deleted[id=V]{then}}{Therefore, we} identified {\obf{several}} improvements that must be carried to the procedure described by \cite{2018A&Atest} when combining Gaia \added[id=A]{angular observations} with radar \added[id=A]{distances}:
\begin{itemize}
\item {\obf{The inclusion of the radar measurements is incomplete in our present work. We account for the range observations but not the Doppler measurements. These type of measurements are rare but can be very useful when very few range observation are available or when the addition of Doppler measurements can help for increasing the time span of the observational data sets as it was the case for (1620) and  (2062).}}
\item {\obf{Addition of MPC observations. In some cases, the prolongation of the time span of the observational data sets can help for improving the quality of the fit and to test the consistency of the present orbits. This next step will be investigated.}}
\item An improvement of the dynamical model. The number of perturbers \added[id=A]{was} \deleted[id=P]{were} increase\added[id=A]{d} from 16 to 353 in order to ensure the \deleted[id=A]{long-term} accuracy of the adjusted orbits \added[id=A]{over an interval longer than the Gaia-DR2 time coverage.}
\item An improvement of the weighting schema. The Gaia systematic error\added[id=A]{s} must be taken into account during the adjustement and not only during the post-fit analysis. This could be done by adding the systematic covariance to the random one when \replaced[id=A]{computing}{creating} the inverse of the weight matrix, or by adding the systematic bias as an explicit parameter to be adjusted during the fit. {\obf{General Relativity will be also included into the asteroid dynamical model.}}
\end{itemize}
Finally, one can expect great perspective\added[id=A]{s} for the future Gaia DR3, in particular concerning mass determination\added[id=A]{s} \replaced[id=V]{, the introduction of the Yarkovsky effect in our dynamical model and tests of General Relavivity with a full consistent planetary and asteroid orbits. 
}{or Yarkovsky effect detection}

\section*{Acknowledgement}
The authors would like to thank the reviewers for their constructive and helpful remarks, in  particular concerning the use of the Gaia systematic error during the post fit analysis and the weighting procedure.  


\bibliographystyle{CDMbasic}
\bibliography{biblio_asteroid_mass}

 \end{document}